\documentclass{aa}  

\usepackage{graphicx}
\usepackage{txfonts}
\usepackage{enumitem}
\usepackage{caption}
\usepackage{subcaption}
\usepackage{float}
\usepackage{fancyvrb}
\usepackage{stfloats}
\usepackage{hyperref}
\hypersetup{colorlinks=true,linkcolor=blue,citecolor=blue,urlcolor=blue}
\begin{document} 

   \title{Algorithms and radiation dynamics for the vicinity of black holes\\
   I. Methods and codes}
   
   \author{Leela Elpida Koutsantoniou}

   \institute{Department of Astrophysics, Astronomy and Mechanics, Faculty of Physics, University of Athens, Panepistimiopolis Zografos, Athens 15784, Greece\newline
   e-mail: \texttt{leelamk@phys.uoa.gr}}

   \date{Received 28 February 2021 / Accepted 8 October 2021}

\abstract{We examine radiation and its effects on accretion disks orbiting astrophysical black holes. These disks are thermally radiating and can be geometrically and optically thin or thick. In this first paper of the series, we discuss the physics and the formulation required for this study. Subsequently, we construct and solve the relativistic radiative transfer equation, or find suitable solutions where that is not possible. We continue by presenting some of the accretion disks we considered for this work. We then describe the families of codes developed in order to study particle trajectories in strong gravity, calculate radiation forces exerted onto the disk material, and generate observation pictures of black hole systems at infinity. Furthermore, we also examine the veracity and accuracy of our work. Finally, we investigate how we can further use our results to estimate the black hole spin and the motion of disk material subjected to these radiation forces.}


   \maketitle


\section{Introduction}

In 2017, one half of the Nobel Prize in Physics was awarded to Barry C. Barish and Kip S. Thorne for the first observation of gravitational waves from merging black holes. In 2019, the Event Horizon Telescope captured the first image of the shadow of the black hole at the core of galaxy M87. And even more recently, one half of the 2020 Nobel Prize in Physics was awarded to Roger Penrose for the discovery that black hole formation is a robust prediction of the general theory of relativity. In the past, black holes may have been considered exotic and mysterious; however, today they are of fundamental importance for astrophysics and are objects of strong ongoing scientific research and investigation. One aspect of that research involves the study of the dynamical effects of radiation in the immediate environment surrounding an astrophysical black hole. One should keep in mind that most known astrophysical black holes are accompanied by accretion disks of a temperature high enough to produce X-rays close to Eddington luminosities in stellar black holes. In fact, before the detection of gravitational waves, this was the only way to detect astrophysical black holes.

In the astrophysical setting of a black hole surrounded by an extended (noncentral) source of radiation (hot accretion disk, corona, and jet), the dynamical effect of radiation, namely radiation pressure, must be taken into account in detail when examining the stability and the evolution of the surrounding matter itself. We can understand this with an analogy from an astrophysically "milder" environment closer to us, namely the Solar System. The dynamical effects of solar radiation were first described by J. H. Poynting \citep{Poynting}. Years later, H. P. Robertson \citep{Robertson} properly explained what is now known as the Poynting--Robertson effect and cleared up misconceptions that had puzzled many great scientists, such as J. Larmor for years. This effect is also called a "drag" because, even though the solar photons are emitted radially outward, they cause the dust grains in orbit absorbing them to slowly brake and infall onto the Sun, the Earth, or any other massive object that is close enough. This force is of 
relativistic origin and is proportional to the absorbing object's azimuthal velocity. A simple nonrelativistic approximative calculation of the radiation force on electrons orbiting a central source of luminosity $L$ is:
\begin{eqnarray}
&&f_{\rm rad}^r =\frac{L \sigma_T}{4\pi r^2 c} , \nonumber \\
&&f_{\rm rad}^\phi\equiv f_{\rm PR}^\phi=-f_{\rm rad}^r \frac{{\rm V}_\phi}{c}=-\frac{L\sigma_T{\rm V}_\phi}{4\pi r^2 c^2} ,
\label{solar}
\end{eqnarray}
where $V_\phi$ is the target's azimuthal velocity, $c$ the speed of light, ${{\sigma }_{T}}$ the Thomson cross section, and $r$ the target's radial distance from the central object.

The goal of this paper is to present the methodology we used and the codes written in order to study the dynamical effects from an extended source of radiation in general relativity, that is way beyond Eqs.~(\ref{solar}). 
The first question we would like to look into is what the magnitude of the radiation created by the hot accretion disk is and whether the effects caused by it could indeed be as negligible as they are often considered to be. We would also like to investigate what the effects of these radiation forces can be on the disk itself and what they could mean for the kinematics, the evolution, and the stability of the disk. We can ponder, for example, if this radiation could perhaps be one of the processes that trigger or regulate the accretion processes. Additionally, we explore the possibly distinct effects that different types of accretion disks can have due to their dissimilar geometrical characteristics, temperature, density gradients, etc. Finally, in this work we present the codes written for our study and show some examples of their results. These include geodesics and trajectories, radiation forces, photograph pictures of black holes surrounded by accretion disks, black hole spin estimation, and radiation induced accretion of material. The results will be presented and discussed in more details in the second part of this work.

Some of the first who studied radiation effects within General Relativity in environments relevant to the ones we examine in the present work are \citealt*{AEL90}, \citealp{ML93,ML96} and \citealp{LM95}. All the aforementioned papers, however, unlike our work, consider central sources of photons, which in some cases are also rotating. A different approach to relativistic radiation effects was followed by \citet{Bini09,Bini11,Bini15}. The work of \citeauthor{SadowskiApJ} (\citeyear{SadowskiApJ,SadowskiPhD,016b}) and \citet{016a} that examine a variety of accretion disks and the effects of radiation and magnetic fields, along with their feedbacks to these systems were also studied and are worth mentioning. The studies of \citet{FW04,FW07}, \citet{FuerstPhD} and \citet{YWF} were also looked into thoroughly. In addition to the main accretion disk environment, we also chose to calculate the radiation force acting on material in outflow regions. We thus looked into original and fundamental accretion, jet and Blandford--Znajek process studies, such as \citet{Bondi1952,BZ,Komissarov1999,Komissarov2001,Livio99,Lee2000,NB04,McKinney2005,KBVK,PNS}, up to more recent general relativistic magnetohydrodynamic (GRMHD) simulation researches on the subjects, such as \citet{Nakamura18,Parfrey19,Park19,Yuan19,EHT,Chatterjee19,Mahlmann20,Konoplya2021,Komissarov2021}.

One other important astrophysical effect related to the complex radiation field around a stellar black hole is the Cosmic Battery \citep{CB98}. This phenomenon operates in all environments including accretion disks, yet its impact is expected to be much more prominent when the central object is more compact, affected by the induced spacetime curvature \citep[see][Table 1]{CB98}. What happens is that the radiation emitted by the hot accretion disk is absorbed by the material itself, exerting on it the aforementioned radiation force. This force, however, is primarily acting upon the material electrons since $f_{e}/f_{p}\sim(m_{p}/m_{e})^{2}$, where ${{f}_{p}}$ and ${{f}_{e}}$ the force on a proton and an electron and ${{m}_{p}}$, ${{m}_{e}}$ their masses respectively. This results in the electrons moving with a different speed than the protons and hence in the generation of a ring current. This current leads consequently to the generation of a poloidal magnetic field that has notable consequences in the structure, equilibrium and evolution of the entire system and its possible outflows. This model was strongly criticized by \citealt*{BKLB}, and was subsequently revisited in \citealt*{CKC06}. Later on, applications were looked into, where notable effects were examined in relevant environments. Additionally, other topics were studied, such as the interaction of the Poynting -- Robertson effect and the Cosmic Battery in X-ray binaries \citep{KCKC} and the repositioning of the inner edge of the accretion disk due to radiation \citep{CP12}.

The primary objective of our research was to study the intensity and effects of radiation in stellar black hole environments and assemble information about systems where the aforementioned Cosmic Battery model could play a mentionable part or have noticeable impacts. Since, nevertheless, the research on supermassive black holes (SMBC) is much more extended and evolving in a much faster pace, particularly the past few years, we also considered that examining studies about the physics of SMBC systems would also be notably constructive, if not necessary. We have thus studied and compared, where possible, our work with the researches of \citet{BLJun05,BLSep05,BLJul06}, \citet{Noble07,Noble11}, \citet{Moscibrodzka09,Moscibrodzka14,Moscibrodzka18} and \citet{Davelaar18}, all of which employ GRMHD.

The present work is a continuation of previous research in \citet{Leela2014} and  \citet{KC2014}, where we first investigated environments with noncentral radiation sources orbiting the central compact object. So far, our work has been the development of ray tracing codes along with custom post-processing algorithms that allow us to redesign and improve the quality and speed of the codes without externally developed (i.e., "black box") components. This was deemed necessary due to the size, duration and complexity of the subject.

The current paper presents and describes a leap forward regarding the quality and effectiveness of our codes as first presented in \citet{KC2014}. Including a new process, we were able to increase the code's resolution by more than a hundredfold, with only doubling the execution time. This allowed us to run a vastly increased number of simulations with finite optical depths at various heights, along and above the equatorial plane, inside and outside the accretion disk, compared to our previous work which involved only targets at the disk's innermost stable circular orbit, hereafter ISCO. Finally, we should mention that the fully ray tracing profile of our codes allows us to look at these objects from very close \citep[see also][]{Davelaar18}. This permits us to study the systems and the incurring radiation forces, as well as obtain images of the black hole and the surrounding accretion disk for radii ranging from $1.25 M$ for a rotating black hole, up to $25 M$ or more, where $M$ is the central black hole mass. This is important because as we see in the results, the closer we travel from the inner edge of the disk toward the black hole, the greater the radiation forces are and the faster their magnitude increases.

In this work, we present in Sect. \ref{S2} the mathematical formulation necessary to set up and use the Kerr metric and the locally nonrotating frames, and the methods to study particle trajectories and radiation effects. In Sect. \ref{S3} we present the various models of disks used in our work and the accompanying physics. In Sect. \ref{S4}, we describe the five different families of codes written for our studies. Finally, in Sect. \ref{S6} we summarize our work and our codes, along with their possible extensions. In addition, we review the significance of this approach and mention the results we discuss in the second part of this work.

\section{Mathematical formulation}
\label{S2}

We assume that the immediate environment around a rotating and accreting black hole, hereafter BH, can be adequately described using the Kerr metric. This suggests that the spacetime is determined by the central compact object that is axisymmetric, uncharged and possibly rotating. We also assume that the presence and motion of test particles does not affect the spacetime form or the stress--energy tensor. We hereafter use the geometrized unit system in which $c=G=1$. We hence measure distances in units of gravitational radii $r_{g}=G M/c^{2}=M$. We also assume the Einstein notation for summation over double indices. Lastly, we denote spacetime components by Greek indices and space components by Latin indices.

\subsection{The Kerr metric}
\label{Kerr metric}
The BH and the spacetime it creates, can be fully described using its mass $M$ and spin parameter $a$. The Kerr metric in Boyer – Lindquist (hereafter BL) coordinates $(t,\phi,r,\theta)$ is given by:
\begin{eqnarray}
{\rm d}{{s}^{2}}&=&
{{g}_{\alpha \beta }} {\rm d}{{x}^{\alpha }}{\rm d}{{x}^{\beta }} \ \nonumber\\
&=&-{{e}^{2\nu }}{\rm d}{{t}^{2}}+{{e}^{2\psi }}{{\left( {\rm d}\phi -\omega {\rm d}t \right)}^{2}}+{{e}^{2{{\mu }_{1}}}}{\rm d}{{r}^{2}}+{{e}^{2{{\mu }_{2}}}}{\rm d}{{\theta }^{2}}\ , \label{metric}
\end{eqnarray}
\noindent where:
\begin{equation}
{{e}^{2\nu }}=\frac{\Sigma \Delta }{A}\ ,\
{{e}^{2\psi }}=\frac{A{{\sin}^{2}}\theta }{\Sigma }\ ,\
{{e}^{2{{\mu }_{1}}}}=\frac{\Sigma }{\Delta }\ ,\
{{e}^{2{{\mu }_{2}}}}=\Sigma\ ,
\end{equation}
\noindent with:
\[
\Delta ={{r}^{2}}-2Mr+{{a}^{2}}\ ,
\]
\[
\Sigma ={{r}^{2}}+{{a}^{2}}{{\cos}^{2}}\theta\ ,
\]
\begin{equation}
A ={{\left( {{r}^{2}}+{{a}^{2}} \right)}^{2}}-{{a}^{2}}\Delta {{\sin}^{2}}\theta\ ,
\end{equation}
\noindent and the spacetime angular velocity is given by:
\begin{equation}
    \omega =-\frac{{{g}_{\phi t}}}{{{g}_{\phi \phi}}}=\frac{2Mra}{A}\ ,\
    \label{S}
\end{equation}
see \citet{B70} and \citet*{B72}.

From the metric (\ref{metric}), we can determine the various characteristic surfaces present around a rotating BH. The event horizon arises from one of the poles of the $g_{rr}$ component and is found at the outermost root of the equation $\Delta = 0$:
\begin{equation}
    r_{\rm evh} = M+\sqrt{M^{2}-a^{2}}.
\end{equation}
The event horizon is thus a sphere of radius $r_{\rm evh} = 2M$ for a nonrotating Schwarzschild BH and $r_{\rm evh} = M$ for a maximally rotating one. The second characteristic surface is the static limit that constitutes the outer boundary surface of the ergosphere and can be found at the point where the $g_{tt}$ component changes sign:
\begin{equation}
    {{r}_{\rm ergo}}=M+\sqrt{{{M}^{2}}-{{a}^{2}}{{\cos }^{2}}\theta }.
\end{equation}

Studying the total energy of circular equatorial orbits \citep[see e.g.,][]{B72}, we can see that there is a limiting case that describes particle orbits of infinite energy per unit rest mass. This is none other than the photon orbit, the innermost circular particle orbit. The radius $r_{ph}$ of this photon ring is given by:
\begin{equation}
    r_{ph}=2M \Bigg\{1+\cos \left[\frac{2}{3}\: \cos^{-1}\left(\mp \frac{a}{M}\right)\right]\Bigg\},
\end{equation}
where the upper sign refers to direct and the lower sign to retrograde orbits. For a Schwarzschild BH with $a=0$, the photon ring radius is $r_{ph}=3M$, while for a maximally rotating BH with $a=M$, we have that $r_{ph}=M$ for the direct and $r_{ph}=4M$ for the retrograde photon orbit.

Finally, another noteworthy set of trajectories are the equatorial circular orbits for massive particles and in particular the ISCO, whose radius is given by:
\begin{equation}
   {{r}_{\rm ISCO}}=M\left[ 3+{{Z}_{2}}\mp \sqrt{\left( 3-{{Z}_{1}} \right)\left( 3+{{Z}_{1}}+2{{Z}_{2}} \right)} \right],
\end{equation}
where:
\[
{{Z}_{1}}=1+{{\left( 1-\frac{{{a}^{2}}}{{{M}^{2}}} \right)}^{{1}/{3}\;}}\left[ {{\left( 1+\frac{a}{M} \right)}^{{1}/{3}\;}}+{{\left( 1-\frac{a}{M} \right)}^{{1}/{3}\;}} \right],
\]
\begin{equation}
    {{Z}_{2}}=\sqrt{\frac{3{{a}^{2}}}{{{M}^{2}}}+Z_{1}^{2}},
\end{equation}
where again the upper sign refers to direct and the lower sign to retrograde orbits. The ISCO starts from a value of ${r_{\rm ISCO}=6M}$ for $a=0$ and for $a=M$, reaches $r_{\rm ISCO}=M$ for a direct orbit. Before moving on, we remark here that the coincidence of the aforementioned characteristic surfaces for a maximally rotating BH at a radius $r=M$ is deceptive and only an artifact of the BL coordinate system. These surfaces and orbits remain separate and distinct for $a=M$, as they differ in radial proper distance \citep[e.g.,][]{B72,Chandra}.

\subsection{Locally nonrotating frames}
The Kerr spacetime is stationary and axisymmetric but the central object rotation introduces complexity in both the physics of the problem and the mathematics required. First of all, the non-diagonality of the metric introduces cumbersome algebraic calculations when rising or lowering indices. In addition, physical difficulties arise when examining locations within the static limit and throughout the ergosphere. This is due to the fact that there cannot be static BL observers at these points, since the $t$ basis vector becomes spacelike.

In order to simplify the calculations and remove various formulation problems inside the ergosphere, we choose to introduce a new set of observers and work in that new frame. The best choice of observers is one where said observers rotate with the spacetime geometry at the point we wish to study. We thus define the locally nonrotating frame (LNRF) or the zero angular momentum observer (ZAMO) at the point in question and describe the requested quantities using their projection on the Minkowskian orthonormal frame of the local observer. If required, we can then easily switch the calculated quantities from the LNRF into the BL frame. We denote quantities calculated in the LNRF by using hats over the component indices (e.g., ${{u}^{{\hat{\alpha }}}}$) and quantities calculated in the BL frame by unhatted indices (e.g., ${{u}^{\alpha }}$). The transformation tensor $e_{\alpha}^{{\hat{\mu}}}$ and $e_{{\hat{\mu}}}^{\alpha}$ between the two frames has nonzero components:
\[
e_{t}^{{\hat{t}}}={{e}^{\nu }} ,\
e_{t}^{{\hat{\phi}}}=-\omega {{e}^{\psi }} ,\
e_{\phi}^{{\hat{\phi}}}={{e}^{\psi }} ,\
e_{r}^{{\hat{r}}}={{e}^{{{\mu }_{1}}}} ,\
e_{\theta }^{{\hat{\theta }}}={{e}^{{{\mu }_{2}}}},
\]
\begin{equation}
e_{{\hat{t}}}^{t}={{e}^{-\nu }} ,\
e_{{\hat{t}}}^{\phi}=\omega {{e}^{-\nu }} ,\
e_{{\hat{\phi}}}^{\phi}={{e}^{-\psi }} ,\
e_{{\hat{r}}}^{r}={{e}^{{{-\mu }_{1}}}} ,\
e_{{\hat{\theta}}}^{\theta}={{e}^{{{-\mu }_{2}}}} .
\label{eQq J}
\end{equation}
Vectors $u^{\alpha}$ and ${{u}^{{\hat{\mu }}}}$, and tensors ${{T}^{\alpha \beta }}$ and ${{T}^{\hat{\mu }\hat{\nu }}}$ are transformed following the equations:
\[
{{u}^{a}}=e_{{\hat{\mu }}}^{\alpha }{{u}^{{\hat{\mu }}}} ,\
{{u}^{{\hat{\mu }}}}=e_{\alpha }^{{\hat{\mu }}}{{u}^{\alpha }} \text{ and}
\]
\begin{equation}
    {{T}^{\alpha \beta }}=e_{{\hat{\mu }}}^{\alpha }e_{{\hat{\nu }}}^{\beta }{{T}^{\hat{\mu }\hat{\nu }}} ,\
    {{T}^{\hat{\mu }\hat{\nu }}}=e_{\alpha }^{{\hat{\mu }}}e_{\beta }^{{\hat{\nu }}}{{T}^{\alpha \beta }} . \label{TabTMN K}
\end{equation}

\subsection{Particle trajectories}
\label{2.3}
In order to study particle trajectories in Kerr spacetime, it is necessary to make full use of the particle's integrals of motion and the respective conserved quantities. Let us assume a particle with rest mass $m$ and four-momentum $p=\left( {{p}_{t}},{{p}_{\phi }},{{p}_{r}},{{p}_{\theta }} \right)$ in geodesic motion around a rotating uncharged BH. This particle has got four conserved quantities: the particle rest mass $m$, the total energy $E=-{{p}_{t}}$, the angular momentum component parallel to the rotation and symmetry axis $L={{p}_{\phi }}$ and the Carter constant $Q={{p}_{\theta }}^{2}+{{\cos }^{2}}\theta \left[ {{a}^{2}}\left( {{\mu }^{2}}-{{p}_{t}}^{2} \right)+{{p}_{\phi }}^{2}/{{\sin }^{2}}\theta  \right]$ \citep{Carter68}. This constant could perhaps be simply explained as a measure of how much a trajectory deviates from the equatorial plane. A particle in geodesic motion that starts in the equatorial plane and has $Q=0$ will remain there indefinitely and a particle moving outside the equatorial plane with $Q>0$ will at some point cross it. Let us note nonetheless, that the magnitude of $Q$ does not relate linearly to the deviation from the equatorial plane motion.

The equations describing the particle motion are:
\[
\Sigma \frac{{\rm d}t}{{\rm d}\lambda }=-a\left( aE{{\sin }^{2}}\theta -L \right)+\left( {{r}^{2}}+{{a}^{2}} \right)\frac{T}{\Delta } ,
\]
\[
\Sigma \frac{{\rm d}\phi }{{\rm d}\lambda }=-\left( aE-\frac{L}{{{\sin }^{2}}\theta } \right)+a\frac{T}{\Delta } ,
\]
\[
\Sigma \frac{{\rm d}r}{{\rm d}\lambda }=\pm \sqrt{{{\mathbb{V}}_{r}}} ,
\]
\begin{equation}
    \Sigma \frac{{\rm d}\theta }{{\rm d}\lambda }=\pm \sqrt{{{\mathbb{V}}_{\theta }}} ,
    \label{LEoM}
\end{equation}
where the effective potentials are given by:
\[
T=\left( {{r}^{2}}+{{a}^{2}} \right)E-aL ,
\]
\[
{{\mathbb{V}}_{r}}={{T}^{2}}-\Delta \left[ {{\mu }^{2}}{{r}^{2}}+{{\left( L-aE \right)}^{2}}+Q \right] ,
\]
\begin{equation}
    {{\mathbb{V}}_{\theta }}=Q-{{\cos }^{2}}\theta \left[ {{a}^{2}}\left( {{\mu }^{2}}-{{E}^{2}} \right)+\frac{{{L}^{2}}}{{{\sin }^{2}}\theta } \right] ,
\end{equation}
and $\lambda$ is an affine parameter for massless particles and $\lambda =\tau /\mu$ for massive particles, with $\tau$ the particle's proper time \citep{B72,Wilkins72}.

The above form of the equations is compact and elegant but hides various problems that appear when one attempts to solve them. The system appears problematic during numerical integration, since the square roots in the latter two Eqs. (\ref{LEoM}) cause the quick accumulation of errors near the turning points. There are various solutions, such as reparameterization, in order to deal with this issue. In our study, we choose to work with the Hamiltonian and transform the above system accordingly. The new system of equations is then as follows:
\[
\frac{{\rm d}t}{{\rm d}\lambda }=-\frac{1}{2\Sigma \Delta }\frac{\partial }{\partial {{p}_{t}}}\left( {{\mathbb{V}}_{r}}+\Delta {{\mathbb{V}}_{\theta }} \right) ,
\]
\[
\frac{{\rm d}\phi }{{\rm d}\lambda }=-\frac{1}{2\Sigma \Delta }\frac{\partial }{\partial {{p}_{\phi }}}\left( {{\mathbb{V}}_{r}}+\Delta {{\mathbb{V}}_{\theta }} \right) ,
\]
\[
\frac{{\rm d}r}{{\rm d}\lambda }=\frac{\Delta }{\Sigma }{{p}_{r}} ,
\]
\[
\frac{{\rm d}\theta }{{\rm d}\lambda }=\frac{1}{\Sigma }{{p}_{\theta }} ,
\]
\[
\frac{{\rm d}{{p}_{t}}}{{\rm d}\lambda }=0 ,
\]
\[
\frac{{\rm d}{{p}_{\phi }}}{{\rm d}\lambda }=0 ,
\]
\[
\frac{{\rm d}{{p}_{r}}}{{\rm d}\lambda }=-{{p}_{r}}^{2}\frac{\partial }{\partial r}\left( \frac{\Delta }{2\Sigma } \right)-{{p}_{\theta }}^{2}\frac{\partial }{\partial r}\left( \frac{1}{2\Sigma } \right)+\frac{\partial }{\partial r}\left( \frac{{{\mathbb{V}}_{r}}+\Delta {{\mathbb{V}}_{\theta }}}{2\Sigma \Delta } \right) ,
\]
\begin{equation}
    \frac{{\rm d}{{p}_{\theta }}}{{\rm d}\lambda }=-{{p}_{r}}^{2}\frac{\partial }{\partial \theta }\left( \frac{\Delta }{2\Sigma } \right)-{{p}_{\theta }}^{2}\frac{\partial }{\partial \theta }\left( \frac{1}{2\Sigma } \right)+\frac{\partial }{\partial \theta }\left( \frac{{{\mathbb{V}}_{r}}+\Delta {{\mathbb{V}}_{\theta }}}{2\Sigma \Delta } \right) . 
\end{equation}
We have therefore transformed the initial four equations of motion into a new system of eight differential equations. The new forms are smooth and do not have poles or other problems throughout their range and can be directly integrated. Let us note that the fifth and sixth of the above equations describe two of the motion's conserved quantities, the conservation of energy and $z$-momentum respectively.

Finally, we define the coordinate angular velocity for a circular equatorial orbit as:
\begin{equation}
    \Omega =\frac{{\rm d}\phi }{{\rm d}t}=\frac{{{u}^{\phi }}}{{{u}^{t}}}=\frac{{{M}^{{1}/{2}\;}}}{{{r}^{{3}/{2}\;}}+a{{M}^{{1}/{2}\;}}},
    \label{T}
\end{equation}
where $u=\left( {{u}^{t}},{{u}^{\phi }},{{u}^{r}},{{u}^{\theta }} \right)$ is the four-velocity of a particle.

\subsection{Radiation and equations of motion}
\label{2.4}

Our main goal in this subsection is to calculate the effects of radiation on the target particle dynamics. Thus, we begin from the formula that relates the target particle position with the acceleration $a^{\alpha}$:
\begin{equation}
    \frac{{{{\rm d}}^{2}}{{x}^{\alpha }}}{{\rm d}{{\tau }^{2}}}+\Gamma _{\mu \nu }^{\alpha }\frac{{\rm d}{{x}^{\mu }}}{{\rm d}\tau }\frac{{\rm d}{{x}^{\nu }}}{{\rm d}\tau }={{a}^{\alpha }} ,
\end{equation}
where $x^{\alpha}$ are the particle position components, $\tau$ the proper time and $\Gamma _{\mu \nu }^{\alpha }$ the Christoffel symbols or connection coefficients \citep{CatalogueofSpacetimes}.

The acceleration in turn, can be given by the relativistic equation of motion:
\begin{equation}
    {{a}^{\alpha }}=\frac{{{f}^{a}}}{m} ,
\end{equation}
where $m$ is the rest mass of the target particle and $f^{\alpha}$ are the nongravitational four-force components. In the present study this is the radiation four-force on the target particle motion.

In order to calculate $f^{\alpha}$, we need to know the flux of the radiation that generates it, as:
\begin{equation}
    {{f}^{\alpha }}=\sigma {{F}^{\alpha }} ,
\end{equation}
where $\sigma$ is the particle cross section for the momentum transfer. The radiation flux four-vector ${{F}^{\alpha }}$ can in turn be calculated using the target particle covariant four-velocity $u_{\mu}$ and the radiation stress – energy tensor $T^{\alpha \beta}$ using the formula:
\begin{equation}
    {{F}^{\alpha }}=h_{\beta }^{\alpha }{{T}^{\mu \beta }}{{u}_{\mu }} ,
\end{equation}
where $h_{\beta }^{\alpha }$ is the projection tensor:
\begin{equation}
    h_{\beta }^{a}=\delta _{\beta }^{\alpha }+{{u}^{\alpha }}{{u}_{\beta }} .
\end{equation}

In order afterwards to acquire the BL radiation stress-energy tensor ${{T}^{\alpha \beta }}$, it is necessary to find the LNRF stress-energy tensor ${{T}^{\hat{\mu }\hat{\nu }}}$ and then use Eq. (\ref{TabTMN K}):
\begin{equation}
    {{T}^{\alpha \beta }}=e_{{\hat{\mu }}}^{\alpha }e_{{\hat{\nu }}}^{\beta }{{T}^{\hat{\mu }\hat{\nu }}} ,
\end{equation}
where the $e_{{\hat{\mu}}}^{\alpha}$ are given by Eq. (\ref{eQq J}).

\begin{figure}
   \centering
   \includegraphics[width=\hsize]{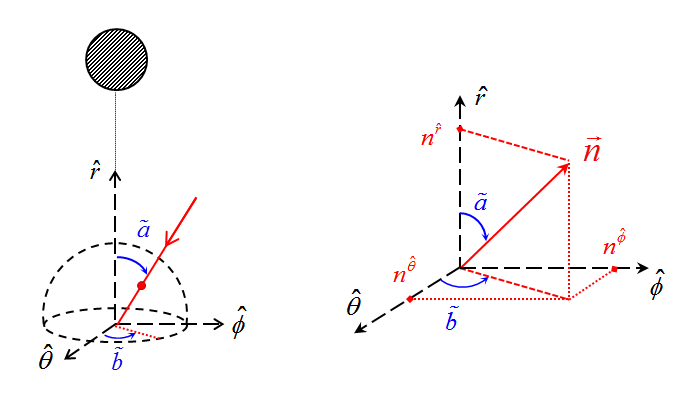}
      \caption{Local sky around the target particle. \textit{Left}: forward in time (the photon moves toward the target). \textit{Right}: moving backwards in time (the photon moves away from the target). For the incoming photon, we define angles $\tilde{a}$ and $\tilde{b}$ similar to the typical polar angle $\theta$ and azimuthal angle $\phi$ of spherical coordinate systems. The disk in stripes at the top left represents the BH event horizon.
      }
         \label{n_decomp}
   \end{figure}

In order to calculate ${{T}^{\hat{\mu }\hat{\nu }}}$ now, we make use of the formula:
\begin{equation}
    {{T}^{\hat{\mu }\hat{\nu }}}=\int{I\left( r, \theta , \tilde{a}, \tilde{b} \right){{n}^{{\hat{\mu }}}}{{n}^{{\hat{\nu }}}}}{\rm d}\tilde{\Omega } ,
\end{equation}
where $I\left( r, \theta , \tilde{a}, \tilde{b} \right)$ is the frequency integrated specific intensity of the radiation, $d\tilde{\Omega }=\sin \tilde{a} \ d\tilde{a} \ d\tilde{b}$ the solid angle element with $\tilde{a}$ and $\tilde{b}$ the related appropriate local angles and ${{n}^{{\hat{\mu }}}}={{{p}^{{\hat{\mu }}}}}/{{{p}^{{\hat{t}}}}}\;$ a unit spacelike vector (Fig. \ref{n_decomp}). Simple calculations can give the ${{n}^{{\hat{\mu }}}}$ vector components as:
\begin{equation}
    {{n}^{{\hat{\phi }}}}=\sin \tilde{a} \sin \tilde{b} , \
    {{n}^{{\hat{r}}}}=\cos \tilde{a} , \
    {{n}^{{\hat{\theta }}}}=\sin \tilde{a} \cos \tilde{b} .
\end{equation}
Intensity $I$ is, as expected, a function of the particle's position in space, since different locations receive different amounts of radiation. From this, we have already excluded the $\phi$ coordinate due to the spacetime axisymmetry. Additionally, $I$ also depends on the angles $\tilde{a}$ and $\tilde{b}$, since different amounts of radiation are received in different orientations of the local sky.

Finally, the frequency integrated specific intensity $I$ is calculated by integrating the specific intensity $I_{\nu}$ across all the contributing frequencies:
\begin{equation}
    I=\int_{\nu }{ {{I}_{\nu }}}  {\rm d}\nu .
\end{equation}
This can be applied to any distribution such as a blackbody, thermal radiation, a single-energy beam of light, or frequency-independent radiation. In an environment where the radiation is emitted by just a surface layer of the source, the calculation of $I$ is relatively easy and it is done with the method described in Sect. \ref{2.6}. On the contrary, in environments where the radiation is emitted by multiple layers or various objects, that method cannot be used. In order to calculate the specific intensity $I_{\nu}$ there, it is required to investigate the radiation transfer process and solve the appropriate equation. This will be addressed in the following subsection.

\subsection{Radiative transfer}
\label{S rad trans 2.5}
In this subsection we look into disks with finite optical depth, where photons are emitted by the material throughout their entire volume. 
The radiation is then regulated by its passage through the disk's absorbing and emitting material. If this material is dense enough, then the ray reaching a point deep inside the disk accumulates a high enough optical depth. This means that the ray cannot have originated from outside the local disk material, but only from inside it. If, on the contrary, the material is of low density or of small quantity, then the ray will travel outside the local material. Later on, the ray can either re-enter the emitting matter further along its path or escape to infinity. In these cases, the incoming ray will only bring in low intensity radiation. We thus attempt here to find a way to calculate the specific intensity $I_{\nu}$ of this radiation. For this reason we look into the radiative transfer equation \citep[see][]{RL}, hereafter RTE,  and the necessary changes required to obtain it in a Lorentz invariant form.

We begin by assuming a thermalized material of number density $n$. This material consists of particles that act as radiation absorbers. We define the absorption coefficient ${{a}_{\nu }}\left( {{\rm cm}^{-1}} \right)$ at frequency $\nu$ as:
\begin{equation}
    {{a}_{\nu}}=n {{\sigma }_{\nu}} ,
    \label{a_nu R}
\end{equation}
where ${{\sigma }_{\nu }}\left( {{\rm cm}^{2}} \right)$ is the absorbing area cross section at a particular frequency. Assuming an initial specific intensity ${{I}_{\nu }}\left( {\rm erg} \ {{\rm cm}^{-2}} {{\rm s}^{-1}} {{\rm ster}^{-1}} {{\rm Hz}^{-1}} \right)$ at frequency $\nu$, the presence of the material's radiation absorbing particles for a propagation length ${\rm d}s$, will cause a decrease in this specific intensity of a propagating light ray given by:
\begin{equation}
    {\rm d}{{I}_{\nu }}=-{{a}_{\nu }}{{I}_{\nu }}{\rm d}s .
\end{equation}
Things are simpler for the emission coefficient ${{j}_{\nu }}\left( {\rm erg} \ {{\rm cm}^{-3}} {{\rm s}^{-1}} {{\rm ster}^{-1}} {{\rm Hz}^{-1}} \right)$. When the light ray propagates for distance ${\rm d}s$, it transverses emitting material of volume ${\rm d}V$ and its specific intensity increases as:
\begin{equation}
    {\rm d}{{I}_{\nu }}={{j}_{\nu }}{\rm d}s .
\end{equation}

The radiative transfer equation combines the above two processes and describes the resulting effects on the light ray's specific intensity as:
\begin{equation}
    \frac{{\rm d}{{I}_{\nu }}}{{\rm d}s}=-{{a}_{\nu }} {{I}_{\nu }}+{{j}_{\nu }} .
\end{equation}
In order to express the solution of the above equation more elegantly, we introduce the concept of the optical depth $\tau_{\nu}$ at frequency $\nu$ that is defined as:
\begin{equation}
    {\rm d}{{\tau }_{\nu }}={{a}_{\nu }}{\rm d}s .
\end{equation}
We can then calculate the optical depth by integrating the above along the path of the light ray:
\begin{equation}
    {{\tau }_{\nu }}\left( s \right)=\int_{{{s}_{0}}}^{s}{{{a}_{\nu }}\left( {{s}\,'} \right) {\rm d}{s}\,'} ,
\end{equation}
where $s_{0}$ is an arbitrarily selected initial point of the scale. We can subsequently restate the RTE as:
\begin{equation}
    \frac{{\rm d}{{I}_{\nu }}}{{\rm d}{{\tau }_{\nu }}}=-{{I}_{\nu }}+\frac{{{j}_{\nu }}}{{{a}_{\nu }}} .
    \label{diff_rte A}
\end{equation}
Integrating this gives the solution to the radiative transfer equation:
\begin{equation}
    {{I}_{\nu }}\left( s \right)={{I}_{\nu }}\left( {{s}_{0}} \right){{e}^{-{{\tau }_{\nu }}}}+\int_{{{s}_{0}}}^{s}{{{j}_{\nu }}\left( {{s}\,'} \right) {{e}^{- \left[ {{\tau }_{\nu }}\left( s \right)-{{\tau }_{\nu }}\left( {{s}\,'} \right) \right]}} {\rm d}{s}\,'} .
    \label{rte}
\end{equation}

   \begin{figure}
   \centering
   \includegraphics[width=\hsize]{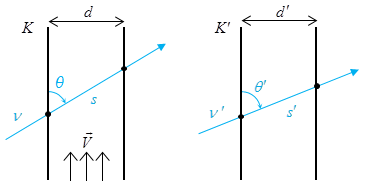}
      \caption{\textit{Left:} emitting matter in the lab frame $K$, moving with velocity $\protect\overrightarrow{V}$ along the vertical axis in a tube of width $d$. A photon of frequency $\nu$ crosses the tube, traveling at an angle $\theta$ from the vertical axis. \textit{Right:} emitting matter in its rest frame ${K}'$ in the tube. The photon with frequency ${\nu}\,'$ now appears to cross the tube at angle ${\theta}\,'$.
      }
      \label{matter_frame}
   \end{figure}

The above magnitudes are not in most cases Lorentz invariant and thus cannot be used in the general solution of the various problems, unless restated in such a form \citep*{MTW}. We begin by considering the phase space number density:
\begin{equation}
    \mathfrak{N}=\frac{N}{\mathcal{V}} ,
\end{equation}
where $N$ is the number of particles under examination and $\mathcal{V}$ the phase space volume they occupy. By taking into account Liouville's theorem in curved spacetime, we have that:
\begin{equation}
    \frac{{\rm d}\mathcal{V}}{{\rm d}\lambda }=0 ,
\end{equation}
where again, $\lambda$ is an affine parameter for massless particles. By combining the above with the conservation of particle number along the world line of the bundle, we obtain that:
\begin{equation}
    \frac{{\rm d}\mathfrak{N}}{{\rm d}\lambda }=0 ,
\end{equation}
which is the collisionless Boltzmann kinetic equation and hence $\mathfrak{N}$ is Lorentz invariant.
The phase space volume is:
\begin{equation}
    {\rm d}\mathcal{V}={{{\rm d}}^{3}}x {{{\rm d}}^{3}}p={\rm d}A\ {\rm d}t\ {{h}^{3}}\ {{\nu }^{2}} {\rm d}\nu  {\rm d}\Omega
\end{equation}
and hence:
\begin{equation}
    \mathfrak{N}=\frac{{\rm d}N}{{{h}^{3}}{{\nu }^{2}}{\rm d}A {\rm d}t {\rm d}\nu  {\rm d}\Omega } .
\end{equation}
Here, $h$ is Planck's constant. Since the specific intensity is defined as:
\begin{equation}
    {{I}_{\nu }}=\frac{h\nu  {\rm d}N}{{\rm d}A {\rm d}t {\rm d}\nu  {\rm d}\Omega } ,
\end{equation}
we can see that:
\begin{equation}
    \mathfrak{N}=\frac{1}{{{h}^{4}}}\frac{{{I}_{\nu }}}{{{\nu }^{3}}}
\end{equation}
and therefore the Lorentz invariant specific intensity ${{\mathcal{I}}_{\nu }}$ is:
\begin{equation}
    {{\mathcal{I}}_{\nu }}=\frac{{{I}_{\nu }}}{{{\nu }^{3}}}=\text{Lorentz invariant} .
    \label{calI_nu B}
\end{equation}
The optical depth, used to count photon fractions, is a scalar quantity and is thus invariant:
\begin{equation}
    \tau =\text{Lorentz invariant} .
\end{equation}
   
In order to find the Lorentz invariant absorption coefficient, we use Fig. \ref{matter_frame}. The tube width $d={d}\,'$ is the same in both the lab and the matter rest frame, since it is perpendicular to the direction of motion. Likewise, the $x$-component of the photon momentum ${{k}_{x}}={{k}_{x}}^{\prime }$, remains unchanged. This subsequently means that $k \sin \theta ={k}\,' \sin {\theta }\,'$ and thus $\nu  \sin \theta ={\nu }\,' \sin {\theta }\,'$. From the Lorentz invariance of the optical depth ${{\tau }_{\nu }}={{a}_{\nu }} s$, we then have:
   \begin{equation}
       {{\mathfrak{a}}_{\nu }}=\nu  {{a}_{\nu }}=\nu { }\,' {{a}_{\nu }}^{\prime }=\text{Lorentz invariant} .
       \label{n_a_nu C}
   \end{equation}
   
Finally, for the emission coefficient we utilize Eqs. (\ref{diff_rte A}), (\ref{calI_nu B}) and (\ref{n_a_nu C}) and conclude that:
\begin{equation}
    {{\mathfrak{j}}_{\nu }}=\frac{{{j}_{\nu }}}{{{\nu }^{2}}}=
    \frac{{{j}_{\nu }}^{\prime }}{{\nu { }\,'{{ }^{2}}}}=\text{Lorentz invariant} .
    \label{j_nu_nu-2 D}
\end{equation}

The Lorentz invariant form of the RTE (\ref{diff_rte A}) will therefore be:
\begin{equation}
    \frac{{\rm d}{{\mathcal{I}}_{\nu }}}{{\rm d}{{\tau }_{\nu }}}=-{{\mathcal{I}}_{\nu }}+\frac{{{\mathfrak{j}}_{\nu }}}{{{\mathfrak{a}}_{\nu }}} .
    \label{dIdtLI E}
\end{equation}
Since for the optical depth, it is ${\rm d}{{\tau }_{\nu }}={{a}_{\nu }} {\rm d}s$, Eq. (\ref{dIdtLI E}) along with (\ref{n_a_nu C}) and (\ref{j_nu_nu-2 D}) gives:
\begin{equation}
    \frac{{\rm d}{{\mathcal{I}}_{\nu }}}{{\rm d}s}=-{{a}_{\nu }}{{\mathcal{I}}_{\nu }}+\frac{{{j}_{\nu }}}{{{\nu }^{3}}} .
    \label{dIn/ds F}
\end{equation}
In order to improve this, we also implicate the path length variation ${{\rm d}s}/{{\rm d}\lambda }$. By using the projection tensor ${{{h}^{\alpha \beta }}={{g}^{\alpha \beta }}+{{u}^{\alpha }}{{u}^{\beta }}}$, we have the photon velocity ${{\left( {{\upsilon }\,'} \right)}^{\alpha }}$ in the fluid frame ${K}'$ as:
\begin{equation}
    {{\left( {{\upsilon }\,'} \right)}^{\alpha }}={{h}^{\alpha \beta }}{{k}_{\beta }}={{k}^{\alpha }}+\left( {{k}_{\beta }}{{u}^{\beta }} \right){{u}^{\alpha }} ,
\end{equation}
where ${{u}^{\alpha }}$ is the fluid four-velocity. By the above, we obtain that:
\begin{equation}
    \frac{{\rm d}s}{{\rm d}\lambda }={{\left. -\left\| {{\left( {{\upsilon }\,'} \right)}^{\alpha }} \right\|  \right|}_{\rm obs}}={{\left. -\sqrt{{{g}_{\alpha \beta }}{{\left( {{\upsilon }\,'} \right)}^{\alpha }}{{\left( {{\upsilon }\,'} \right)}^{\beta }}} \right|}_{\rm obs}}={{\left. -{{k}_{\beta }}{{u}^{\beta }} \right|}_{\rm obs}}
\end{equation}
and for the frequency ratio, it is:
\begin{equation}
    \frac{\nu }{{{\nu }\,'}}=\frac{{{\left. {{k}_{\beta }}{{u}^{\beta }} \right|}_{\rm obs}}}{{{\left. {{k}_{\alpha }}{{u}^{\alpha }} \right|}_{\lambda }}} ,
    \label{nu/nu' G}
\end{equation}
\citep[see also][]{YWF}. Quantities with an accent, such as $\nu { }\,'$ above, are henceforth measured in the local rest frame. Combining these, we have that:
\begin{equation}
    \frac{{\rm d}s}{{\rm d}\lambda }={{\left. -{{k}_{\alpha }}{{u}^{\alpha }} \right|}_{\lambda }}\ \frac{\nu }{{{\nu }_{0}}} .
    \label{dcalI/dl H}
\end{equation}
Eq. (\ref{dIn/ds F}) combined with (\ref{n_a_nu C}), (\ref{j_nu_nu-2 D}) and (\ref{dcalI/dl H}) gives the differential form of the invariant RTE, the general relativistic radiative transfer equation (GRRTE) equation as:
\begin{equation}
    \frac{{\rm d}{{\mathcal{I}}_{\nu }}}{{\rm d}\lambda }=-{{\left. {{k}_{\alpha }}{{u}^{\alpha }} \right|}_{\lambda }}\left( -{{a}_{\nu }}^{\prime }{{\mathcal{I}}_{\nu }}+\frac{{{j}_{\nu }}^{\prime }}{\nu {{{{ }\,'}}^{3}}} \right) .
\end{equation}
Integration of the above gives the solution for the Lorentz invariant specific intensity:
\begin{eqnarray}
    {{\mathcal{I}}_{\nu }}\left( \lambda  \right)&=&{{\mathcal{I}}_{\nu }}\left( {{\lambda }_{0}} \right){{e}^{\int_{{{\lambda }_{0}}}^{\lambda }{{{a}_{\nu }}^{\prime }\left( \zeta  \right) {{\left. {{k}_{\alpha }}{{u}^{\alpha }} \right|}_{\zeta }}{\rm d}\zeta }}}\nonumber\\ &-&\int_{{{\lambda }_{0}}}^{\lambda }{\frac{{{j}_{\nu }}^{\prime }\left( \xi  \right)}{\nu { }\,'{{ }^{3}}}{{e}^{\int_{\xi }^{\lambda }{{{a}_{\nu }}^{\prime }\left( \zeta  \right) {{\left. {{k}_{\alpha }}{{u}^{\alpha }} \right|}_{\zeta }}{\rm d}\zeta }}}}{{\left. {{k}_{\alpha }}{{u}^{\alpha }} \right|}_{\xi }}{\rm d}\xi.
    \label{I}
\end{eqnarray}
The optical depth can be calculated as:
\begin{equation}
    {{\tau }_{\nu }}\left( \lambda  \right)=-\int_{{{\lambda }_{0}}}^{\lambda }{{{a}_{\nu }}^{\prime }\left( \zeta  \right) {{\left. {{k}_{\alpha }}{{u}^{\alpha }} \right|}_{\zeta }}{\rm d}\zeta }
\end{equation}
and equation (\ref{I}) can then be rewritten as:
\begin{equation}
    {{\mathcal{I}}_{\nu }}\left( \lambda  \right)={{\mathcal{I}}_{\nu }}\left( {{\lambda }_{0}} \right){{e}^{-{{\tau }_{\nu }}\left( \lambda  \right)}}-\int_{{{\lambda }_{0}}}^{\lambda }{\frac{{{j}_{\nu }}^{\prime }\left( \xi  \right)}{\nu { }\,'{{ }^{3}}}{{e}^{-\left[ {{\tau }_{\nu }}\left( \lambda  \right)-{{\tau }_{\nu }}\left( \xi  \right) \right]}}}{{\left. {{k}_{\alpha }}{{u}^{\alpha }} \right|}_{\xi }}{\rm d}\xi .
\end{equation}

\subsection{Intensity of single emission source radiation}
\label{2.6}
In this subsection we describe the way to estimate the radiation received by a target, when said radiation is emitted by a single emission source. This means that the photons are emitted by a skin surface of the accretion disk, henceforth AD, and do not traverse any of its material. This happens in the case where the disk is totally optically thick. Various parts of this procedure have been studied in the literature: \citet*{AEL90} studied radiation emitted by a central nonrotating star in Schwarzschild spacetime. \citet{ML96} also studied the environment around emitting stars and expanded this work by examining nonrotating and rotating masses and radiating sources. We have also studied this subject in the previous work \citet{KC2014} examining fewer examples of totally opaque disks and a single observer position for each model-BH spin set. Here, apart from expanding into semi-opaque disks discussed later on, we expand our analysis to more disk models and instead of having a single observer at the ISCO of each model-spin set, we fill the entire region of the system with a large amount of observers in different locations.

As we saw previously in Eq. (\ref{calI_nu B}), the Lorentz invariant specific intensity is ${{\mathcal{I}}_{\nu }}={{{I}_{\nu }}}/{{{\nu }^{3}}}\;$ and thus for the frequency integrated specific intensity it is:
\begin{equation}
    \frac{{{I}_{1}}}{\nu _{1}^{4}}=\frac{{{I}_{2}}}{\nu _{2}^{4}}=\text{Lorentz invariant},
\end{equation}
for any two random points $P_1$ and $P_2$. From this, we have that for the emitted and the received frequency integrated specific intensity of a photon, ${{I}_{\rm em}}$ and ${{I}_{\rm rec}}$ respectively, it is:
\begin{equation}
    {{I}_{\rm rec}}={{\left( \frac{{{\nu }_{\rm rec}}}{{{\nu }_{\rm em}}} \right)}^{4}}{{I}_{\rm em}} .
\end{equation}
We note here that the frequency fraction ${{{\nu }_{\rm rec}}}/{{{\nu }_{\rm em}}}$ that appears above does not depend on the frequencies involved, but only on the spacetime and the photon's emission angle. This frequency fraction includes the effects of three different phenomena caused mainly by the spacetime properties. Firstly, it includes the effects of gravitational time dilation, which appear both in Schwarzschild and Kerr spacetimes. It also includes the frame dragging frequency shift due to the spacetime's differential rotation that appears only in a Kerr spacetime. Finally, it includes the Doppler shift caused by the motion of the source's emitting surface. This can exist in both Schwarzschild and Kerr spacetimes.

The gravitational time dilation causes a frequency shift for the received frequency ${{\nu }_{\rm rec}}$ given by:
\begin{equation}
    {{\nu }_{\rm rec}}={{\left( \frac{{{g}_{tt, {\rm em}}}}{{{g}_{tt, {\rm rec}}}} \right)}^{{1}/{2}\;}}{{\nu }_{\rm em}} ,
\end{equation}
where ${{\nu }_{\rm em}}$ the emitted frequency.

   \begin{figure}
   \centering
   \includegraphics[width=\hsize]{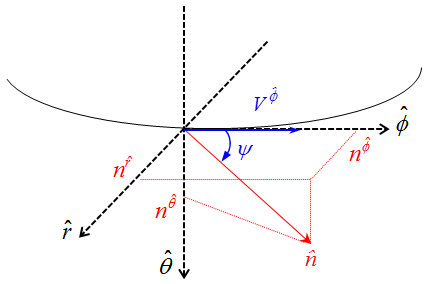}
      \caption{Emission of a photon from the hot AD. The photon is emitted along $\hat{n}$ from an element of matter moving with velocity ${{V}^{{\hat{\phi }}}}$. The photon is emitted at an angle $\psi $ relative to the accreting material motion.
      }
         \label{photon_emission}
   \end{figure}

Assuming the photon source moves azimuthally with negligible radial and poloidal velocity components, the Doppler shift due to the emitting surface motion then introduces a change in frequency:
\begin{equation}
    {{\nu }_{\rm rec}}=\frac{1}{\gamma \left( 1-{{V}^{{\hat{\phi }}}} \cos \psi  \right)}{{\nu }_{\rm em}} ,
\end{equation}
where $V ={{V}^{{\hat{\phi }}}}$ the source three-velocity here, $\gamma ={{\left( 1-{{V}^{2}} \right)}^{-{1}/{2}\;}}$ the emitting material Lorentz factor and $\psi $ the angle between the emitting matter velocity and the photon emission direction (Fig. \ref{photon_emission}). Let us mark that both the $\gamma $ factor and the $\psi $ angle are measured in the ZAMO frame at the point of emission.

\begin{figure*}
   \centering
   \includegraphics[width=\hsize]{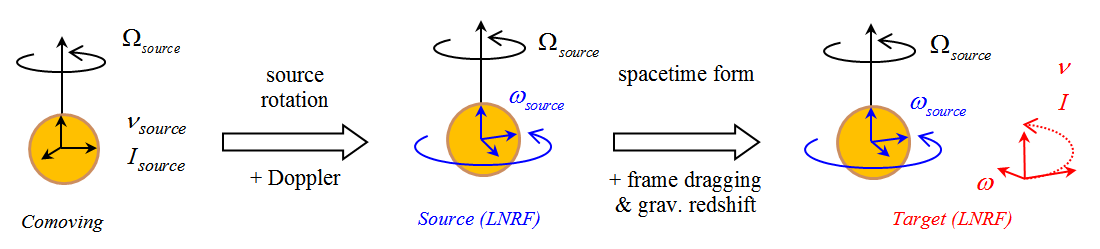}
      \caption{Schematic of the frequency changes. A Doppler shift is used to move from the frame comoving with the emitting surface (\textit{left}) to the LNRF at the radius of the surface (\textit{middle}), taking into account the emission source rotation. Then we move to the receiving particle frame (\textit{right}), accounting for two more changes in frequency, the gravitational time dilation due to the change of radial distance from the source to the target and the different effects of frame dragging, again because of the change in radial distance.
      }
         \label{system_changes}
   \end{figure*}

Concluding the necessary transformations, we have the factor required for the implementation of the frame dragging effects, which is:
\begin{equation}
    {{\nu }_{\rm rec}}=\frac{1+{{\omega }_{\rm rec}}\: {{{k}_{\phi }}}/{{{k}_{t}}}\;}{1+{{\omega }_{\rm em}}\: {{{k}_{\phi }}}/{{{k}_{t}}}\;} {{\nu }_{\rm em}} ,
\end{equation}
where ${{k}_{a}}$ are the photon covariant four-momentum components, which are also conserved quantities. The ratio ${{{k}_{\phi }}}/{{{k}_{t}}}$ depends only on the direction of the photon emission and from the previous statement is also a conserved quantity. Combining the above, we have for the received frequency that:
\begin{equation}
    {{\nu }_{\rm rec}}={{\left( \frac{{{g}_{tt, {\rm em}}}}{{{g}_{tt, {\rm rec}}}} \right)}^{{1}/{2}\;}}\frac{1+{{\omega }_{\rm rec}}\: {{{k}_{\phi }}}/{{{k}_{t}}}\;}{1+{{\omega }_{\rm em}}\: {{{k}_{\phi }}}/{{{k}_{t}}}\;} \frac{1}{\gamma \left( 1-{{V}^{{\hat{\phi }}}} \cos \psi  \right)}{{\nu }_{\rm em}}
\end{equation}
and therefore for the frequency integrated specific intensity:
\begin{equation}
    {{I}_{\rm rec}}={{\left( \frac{{{g}_{tt, {\rm em}}}}{{{g}_{tt, {\rm rec}}}} \right)}^{2}}{{\left( \frac{1+{{\omega }_{\rm rec}}\: {{{k}_{\phi }}}/{{{k}_{t}}}\;}{1+{{\omega }_{\rm em}}\: {{{k}_{\phi }}}/{{{k}_{t}}}\;} \right)}^{4}} \frac{1}{{{\gamma }^{4}}{{\left( 1-{{V}^{{\hat{\phi }}}} \cos \psi  \right)}^{4}}} {{I}_{\rm em}} .
\end{equation}
In Fig. \ref{system_changes}, we can see a breakdown of the process described above, where for visual simplicity we assume that the emission of photons is done by a central object.

\section{Accretion tori}
\label{S3}
In this section we present the accretion tori we used for our codes. Some of the tori are optically thick while others are stratified and semi-opaque. In general, we examine tori of assorted geometrical shapes and diverse density profiles. This way we can better cover for example the various stages of matter infall into the BH and the different stages of disk evolution in X-ray binaries. For the physics, evolution, magnetic fields role and various other phenomena present in the intriguing X-ray binary environments, one could consult \citet{Esin97,Verbunt99,Bildsten01,Fender02,Haggard04,Mueller04,DGK,Meyer09,vanHaaften12,Heinke13}.

A broad and thorough study of compact objects, accretion disks and the many physical phenomena observable in such environments was presented by \citet*{BABHKL}. In order for us to design and choose acceptable disk models with a good balance between model quality and computational time, we looked into \citet{SS73}, \citet{Cunningham75,Cunningham76}, \citet{A78,A88,A96}, \citet{K78}, \citet{NY94,NY95}, \citet{Lasota99}, \citet{Igumenshchev}, \citet{NIA,NSPK}, \citet{N08}, \citet{SadowskiApJ}, \citet{Noble11}, \citet{PSKN}, \citet{FuerstPhD} and references therein, \cite{SadowskiPhD} and references therein. We have thus constructed two different AD groups, one with optically thick disks and one with semi-opaque or translucent disks. The main repercussions of the disk's thermal radiation, along with its impact onto the disk's geometry, vertical height and optical thickness were described and studied in \citet{TP75,InoueHoshi,Takahashi95,Beloborodov98,Beloborodov99,Beloborodov01,DynamicProcesses,AF13}. For the finer points that differentiate the optically and geometrically thin and thick disks and segregate the categories, one could look for example into \citet{Artemova}, \citet{LREF} and \citet{Dubus}.

We note here, that the ADs we considered for this work are just a sample for the study of the most commonly considered models. Many other frequently used AD models, such as Novikov--Thorne disks \citep{Novikov-Thorne} for example, could also be implemented and studied using our codes, should the need arise. The reason why we chose to consider more simplistic perhaps disk models than the latest GRMHD researches (e.g \citealt{Chatterjee19,Mahlmann20}), is because of the number of code executions this work required. This results in any model improvement, such as a more realistic, a nonaxisymmetric or a time-variant AD, greatly increasing the total execution time.

The first group of disks we examine is used to represent tori that increase their density abruptly and very close to their outer surface. They are mostly rotationally supported and totally optically thick. This practically means that there is no reason to solve the GRRTE for these tori and instead another methods of calculation must be used. We use these models to describe physical tori that are either cold or compact, or both. An interesting case they could also be used to describe, is transient stages of the X-ray binary systems \citep[see e.g.,][]{FECS}. During the quiescent stages of these systems, their ADs tend to be cooler and at a larger distance from the compact object \citep[e.g.,][]{Esin97,N08}. They subsequently remain in a similar state for an indefinite amount of time. At some point later on, they start increasing their temperature and swelling up while reducing their density and density gradient. From that point on, we can no longer describe them using these opaque models and must instead employ semi-opaque tori models. Additional and detailed information about the evolution and stages of X-ray binaries can be found in \citet{Tauris00}, \citet{Pods02} and \cite{Chen16}.

The second group of tori, the semi-opaque ones, describes more common and familiar perhaps cases of ADs. There is a measurable density and temperature gradient. For these cases, we must use a ray tracing process and solve the GRRTE along the photon trajectory. This way, we calculate how much radiation is produced by the hot material in every step and how much of this is absorbed away by it. Depending on the direction and angle of motion of the traveling photon, it can at times be absorbed by the disk material and, at other times, it can traverse part of the disk without it being absorbed. This means that at some points the disk is optically thick and at other times optically thin, hence the name semi-opaque. In these cases, we observe effects such as transparency and limb darkening (Fig. \ref{2x2 pic}, bottom left \& right).

All the tori we mention here can rotate in various ways. In our program executions we assume that the disk material can rotate circularly ($u^r=0$) with the typical coordinate angular velocity $\Omega$ (Eq. \ref{T}) or with more detailed profiles such as the one in Eq. (\ref{U}) below. Additionally, we examine cases where the disk material follows inspiral motion profiles ($u^r\neq0$) attempting to mimic the SANE (Standard And Normal Evolution) and MAD (Magnetically Arrested Disk) models \citep{NSPK,PSKN,NIA}. In each execution run, our codes give results 
for all of the aforementioned different velocity profiles for the AD material.

We should also mention here that although in many of the AD simulations and studies it is assumed that the material is in a stationary condition, geometrically at least, in reality it is far from that. There are increased amounts of turbulence, instabilities \citep[see e.g.,][]{TNM,MTB,NSPK} and other phenomena taking place, often in smaller scales, that are at times ignored. One such phenomenon is the flow and diffusion of angular momentum throughout the different disk sectors. The main cause of this diffusion is considered to be the material viscosity and is treated in various ways. One of the best known and more frequently used methods is the $\alpha$-viscosity approach \citep{SS73,A88}. This method, nevertheless, cannot be applied to all disk models, such as non $\alpha$-disks, where other solutions must be found.

Another important phenomenon that is known and generally mentioned in such works but often ultimately ignored, is the existence of magnetic fields. The presence of magnetic fields usually gives rise to very important phenomena, such as the Blandford–Znajek process \citep{BZ,Livio99,Komissarov2001,McKinney2005,KBVK,PNS}, that affect the structure of the disk and its stability, and determine its evolution. In theory, the Blandord-Znajek process is, along with the Penrose process \citep{Penrose71}, one of the two most promising phenomena responsible for launching astrophysical jets. The presence of substantial poloidal magnetic field makes the extraction of spin energy and angular momentum from a rotating BH possible. This happens due to the escape of angular momentum from the rotating magnetosphere inside the ergosphere. This mechanism can accurately describe the formation and ejection of jets from spinning SMBHs and is also considered to play a pivotal part in gamma-ray bursts.

An additional noteworthy effect brought on by the existence of magnetic fields and very important for the AD dynamics, is the generation of magnetorotational instabilities (MRI). This strong fluid instability occurs when a conductive AD is situated in a magnetic field and is rotating differentially, with its inner regions rotating faster than its outer regions. The freely moving charges of the material are subjected to the Lorentz force, due to the presence of the magnetic field. Any fluid element deviating even to a small extent from circular motion, has its trajectory further destabilized by a force increasing proportionally to the displacement from the circular orbit. This causes the disk to become unstable and consequently turbulent. The MRI can therefore have important consequences, particularly on the distribution and flow of angular momentum throughout the AD and its diffusion toward the outer layers and components \citep[see][]{BHa91,BHd92,BHb91,BHc92,HGB,BH98,KrolikB,KrolikP,TSKS,PCP}. Other phenomena the MRI is also expected to influence is the formation of active galactic nuclei \citep{KrolikB}, the production of X-rays in compact object systems \citep{Blaes04}, as well as gamma-ray bursts \citep{Wheeler04}.

\subsection{Optically thick accretion tori}
\label{3.1}
In this subsection we describe the models we used for opaque tori. We built some of these tori by assuming simplistic disk cross section shapes, such as polygons. These tori can be viewed either as toy models or as initial condition "snapshot" states. One could then go on to study the evolution of these tori taking into account the presence of radiation effects. Some of the other models we considered are more complex. We built those models self-consistently by assuming that the material of the disk is supported and kept in place by its rotation.

Optically thick ADs are generally expected to be geometrically thin \citep[e.g.,][]{SS73}, even though that is not always a canon, as described in the aforementioned accretion disk studies. This is caused by the “inefficiency” of the radiation: since the disk material is opaque, the radiation transmitted by its hot components cannot reach other, more distant parts of the disk. This results in each local material component to have a significantly lesser “inflating” radiation pressure element than a “deflating” gravitational force element. The result is an AD of smaller geometrical thickness and a much larger pressure gradient, specifically close to its outer surface. We, however, investigate both the cases of geometrically thin and thick opaque ADs.

A matter of particular importance is the surface temperature distribution we assume for these tori. For our calculations, we considered two separate cases, an isothermal disk and a disk whose temperature follows $T\propto {{r}^{-{3}/{4}}}$ \citep{SS73}. The first case, albeit unnatural, is the simplest possible one could imagine and is thus perhaps easier to understand and effortlessly anticipate certain results.

The second case is to assume that the disk temperature distribution is caused by the material accreted onto the central compact object. If we assume that the object's luminosity is equal to the Eddington luminosity ${{L}_{\rm Edd}}={4\pi G\mathcal{M}{{m}_{p}}c}/{{{\sigma }_{T}}}$, then we have an Eddington accretion rate:
\begin{equation}
    {{\dot{\mathcal{M}}}_{\rm Edd}}=\frac{4\pi G\mathcal{M}{{m}_{p}}}{c {{\sigma }_{T}}} ,
\end{equation}
where $G$ is the gravitational constant, $\mathcal{M}$ the disk mass, and ${{m}_{p}}$ the mass of the proton. Assuming a large enough amount of scatterings, the disk material can be adequately described by blackbody radiation. Then, its temperature will be given by:
\begin{equation}
    T={{\left( \frac{3G\dot{\mathcal{M}}\mathcal{M}}{8\pi \sigma_{\rm SB} } \right)}^{{1}/{4}}}{{r}^{{-3}/{4}}} ,
\end{equation}
where $\sigma_{\rm SB} $ is the Stefan – Boltzmann constant \citep[see][]{Longair}.

We notice here, nevertheless, the problem that arises if we simplistically hypothesize the above. If we assume that the accretion luminosity is equal to the Eddington luminosity, then the disk cannot be geometrically thin. This is because, as the accretion luminosity increases, the radiation pressure exerted onto the material keeps getting larger and finally comparable to local gravitational forces. As this happens, the disk keeps inflating by gaining height and width and thus gradually turning into a geometrically thick and optically thin torus \citep[e.g.,][]{TP75}. The easiest way to bypass such problems is to assume that the accreting object radiates only a fraction $\epsilon $ of the Eddington luminosity:
\begin{equation}
    {{\dot{\mathcal{M}}}_{\rm acc}}=\epsilon  {{\dot{\mathcal{M}}}_{\rm Edd}}
\end{equation}
and thus, it is:
\begin{equation}
    T={{\left( \frac{3\epsilon G\dot{\mathcal{M}}\mathcal{M}}{8\pi \sigma_{\rm SB} } \right)}^{{1}/{4}}}{{r}^{{-3}/{4}}} .
\end{equation}

Finally, after having picked any of the disk temperature profiles, we can have its effect on the emitted photon from:
\begin{equation}
    {{I}_{\rm em}}=\frac{{\sigma_{\rm SB} }}{\pi }T{{\left( {{r}_{\rm em}} \right)}^{4}} \ .
\end{equation}

In our work, we have considered so far six different models for optically thick accretion tori. Specified by their given names, we have the models band, disk, slab, wedge, torus and opaque rotationally supported torus:
\begin{enumerate}[label=(\alph*)]
\item Band (toy, snapshot model): a cylindrical surface of half-height $h$ (from its highest point to the equatorial plane) at the distance of the respective ISCO for the selected spin parameter (Fig. \ref{2D_disks}). We can freely choose the half-height $h$ without restrictions and we can also adjust the cylinder radius. We use this mode when we wish to study for example the radiation effects on only the innermost surface of an AD.
\item Disk (toy, snapshot model): an infinitesimally thin disk at the equatorial plane (Fig. \ref{2D_disks}). Its inner radius is at the distance of the ISCO and its outer radius at any multiple of this distance. The innermost and outermost radius of the disk can easily be adjusted.
   \begin{figure}
   \centering
   \includegraphics[width=\hsize]{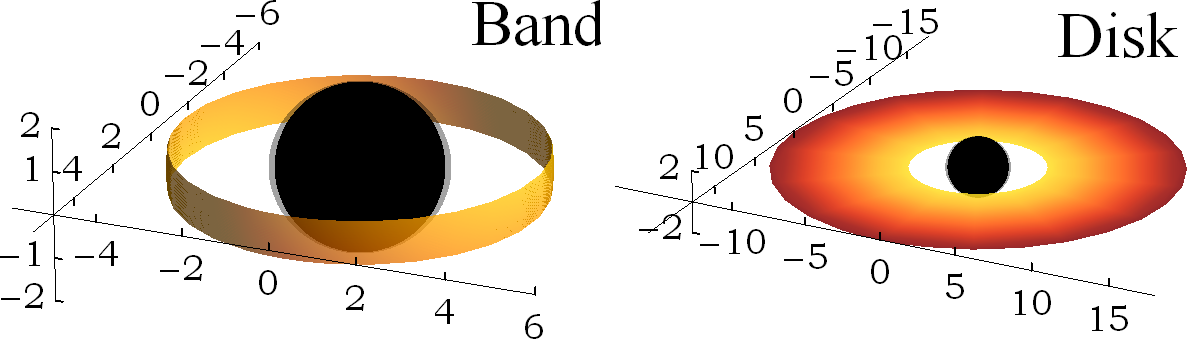}
      \caption{ADs "Band" (\textit{left}) and "Disk" (\textit{right}) constituted by two-dimensional surfaces (Sect. \ref{3.1}a and b respectively). In the center of each image is in black the BH event horizon and around it in gray, its ergosphere. We see the AD in the yellow (hotter) and red tones (colder).
      }
         \label{2D_disks}
   \end{figure}
\item Slab (toy, snapshot model): a disk of half-height $h$ from the radius of the ISCO to a distance of three times the ISCO radius. The cross section of the disk is a rectangle (Fig. \ref{opaqdk}). The half-height $h$ can be freely chosen without restrictions. The innermost and outermost radius of the AD can also be adjusted.\label{slabc}
\item Wedge (toy, snapshot model): a disk whose cross section is an isosceles trapezoid and is centered above and below the equatorial plane. Its inner radius is equal to the radius of the ISCO and its outer radius to three times that. We construct the disk in such a way that an angle with its vertex at the BH (the origin) extending outward, reaches the ISCO cylinder intersecting a ring of half-height $h$. The angle sides continue extending outward in the same direction until crossing the outer edge of the disk (Fig. \ref{opaqdk}). The disk half-height $h$, as well as its inner and outer radius can easily be modified.\label{wedged}
\item Torus (toy, snapshot model): a disk with a circular cross section. The center of the circle is at coordinates ${\left( r, \theta  \right)=\left( 2 {{r}_{\rm ISCO}}, {\pi }/{2}\; \right)}$ and its radius is equal to the ISCO radius. The disk inner edge is therefore at $r={{r}_{\rm ISCO}}$ and the outer edge at $r=3 {{r}_{\rm ISCO}}$. The cross section center and radius of the disk can be adjusted at will (Fig. \ref{opaqdk}).\label{toruse}
   \begin{figure}
   \centering
   \includegraphics[height=0.9525 \textheight,keepaspectratio]{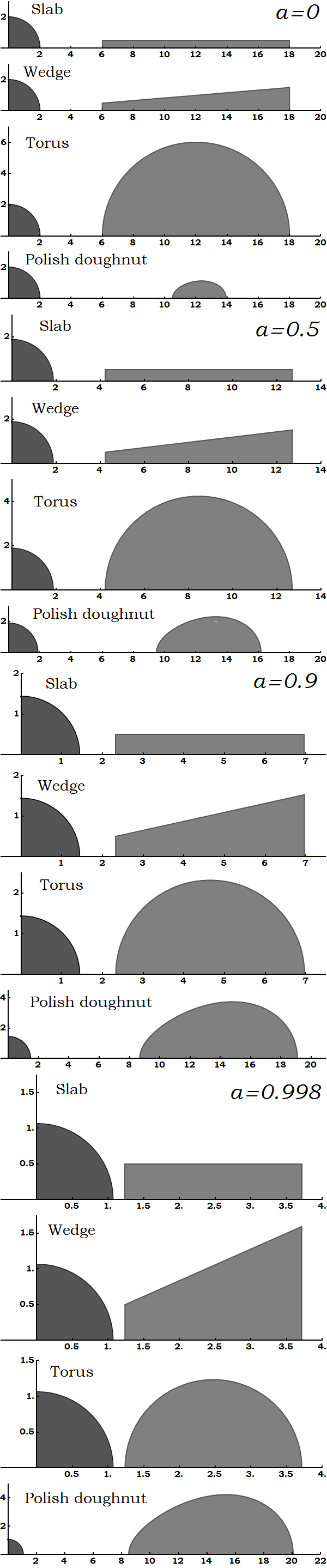}
      \caption{Opaque ADs cross sections for the spins studied \\(Sect. \ref{3.1}c--f).
      }
         \label{opaqdk}
   \end{figure}
   
\item Opaque rotationally supported torus (ORST -- self-consistent model): a rotationally supported torus. This disk is one of the more complex cases considered for optically thick disk examples. The disk we consider here is stationary and axisymmetric and has its rotation axis aligned with the rotation axis of the BH. In our work we have assumed that the two angular velocity vectors are collinear, but it is simple to consider the opposite case in order to study retrograde disks. We then assume that the disk acceleration is what creates this setup and specifies its shape. The acceleration along the particle trajectory is given by:
\begin{equation}
    {{a}^{\alpha }}=\frac{\rm{D}{{u}^{\alpha }}}{\rm{d}\tau }={{u}^{\alpha }}_{;\,\beta }{{u}^{\beta }}={{u}^{\alpha }}_{,\,\beta }{{u}^{\beta }}+\Gamma _{\beta \rho }^{\alpha }{{u}^{\beta }}{{u}^{\rho }} ,
    \label{P}
\end{equation}
where $\Gamma _{\beta \rho }^{\alpha }$ are the Christoffel symbols which we can calculate from the metric (\ref{metric}), using the formula:
\begin{equation}
    \Gamma _{\kappa \lambda }^{\alpha }=\frac{1}{2}{{g}^{\alpha \mu }}\left( {{g}_{\mu \kappa ,\lambda }}+{{g}_{\mu \lambda ,\kappa }}-{{g}_{\kappa \lambda ,\mu }} \right) 
\end{equation}
(more details can be found in \citealt*{AMP}). The torus we have assumed here, like many of the tori in works of this type, has negligible radial and poloidal velocity components, so (\ref{P}) can be slightly simplified. We then get for the acceleration components that:
\begin{equation}
{{a}^{t}}\equiv 0,
\label{LMNO}
\end{equation}
\[{{a}^{\phi }}\equiv 0,\]
\[{{a}^{r}}=-\frac{\Delta }{\Sigma }\left[ M\frac{\Sigma -2{{r}^{2}}}{{{\Sigma }^{2}}}{{\left( {{u}^{t}}-a  {{u}^{\phi }} {{\sin }^{2}}\theta \right)}^{2}}+r  {{\left( {{u}^{\phi }} \right)}^{2}} {{\sin }^{2}}\theta \right],\]
\[{{a}^{\theta }}=-\frac{\sin \theta  \cos \theta }{\Sigma }\left\{ \frac{2Mr}{{{\Sigma }^{2}}}{{\left[ a {{u}^{t}}-\left( {{r}^{2}}+{{a}^{2}} \right){{u}^{\phi }} \right]}^{2}}+\Delta {{\left( {{u}^{\phi }} \right)}^{2}} \right\} . \]
The first two of the above equations are in accordance with our initial assumption that the disk is stationary and axisymmetric respectively. The last two equations respectively are what can give the surface of constant acceleration, the isobaric surfaces through:

   \begin{figure}
   \centering
   \includegraphics[width=\hsize]{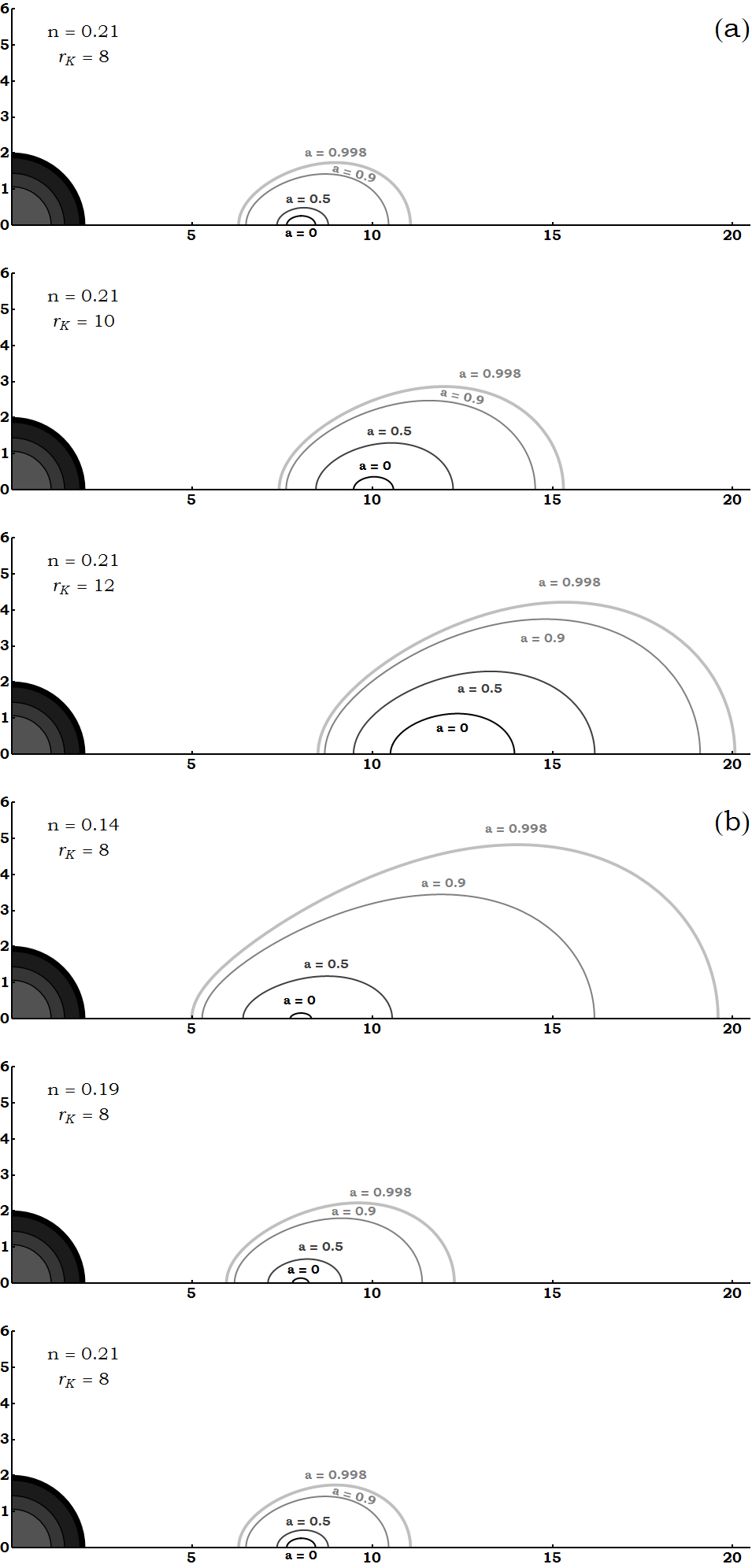}
      \caption{ORST cross sections (Sect. \ref{3.1}f). In (a) we can see the effects caused by the change of the radius of Keplerian rotation speed $r_{K}$. In (b) we see the tori shapes and sizes for various values of the parameter $n_{p}$.
      }
         \label{orsts}
   \end{figure}
   
\begin{equation}
    {{a}_{\alpha }}{{u}^{\alpha }}=0 .
\end{equation}
Using the above equations, this gives:
\begin{equation}
    {{a}_{r}}{{u}^{r}}+{{a}_{\theta }}{{u}^{\theta }}=\frac{\Sigma }{\Delta }{{a}^{r}}{{u}^{r}}+\Sigma {{a}^{\theta }}{{u}^{\theta }}=0 .
\end{equation}
This leads to the pair of differential equations:
\[\frac{{\rm d}r}{{\rm d}\xi }=\frac{Y}{\sqrt{{{Y}^{2}}+\Delta  {{X}^{2}}}} , \]
\begin{equation}
    \frac{{\rm d}\theta }{{\rm d}\xi }=-\frac{X}{\sqrt{{{Y}^{2}}+\Delta  {{X}^{2}}}} ,
\end{equation}
where:
\[X=M\frac{\Sigma -2{{r}^{2}}}{{{\Sigma }^{2}}}{{\left( {{\Omega }^{-1}}-a {{\sin }^{2}}\theta  \right)}^{2}}+r {{\sin }^{2}}\theta, \]
\begin{equation}
   Y=\sin \theta  \cos \theta \left[ \frac{2Mr}{{{\Sigma }^{2}}}{{\left( a {{\Omega }^{-1}}-{{r}^{2}}-{{a}^{2}} \right)}^{2}}+\Delta  \right]
\end{equation}
and $\Omega ={{{u}^{\phi }}}/{{{u}^{t}}}\;$ is, as before, the material's angular velocity. The last information necessary to solve the above and have the resulting torus is its inner edge at the equator. We find this by solving the equation for marginal stability orbits:
\[2aM{{\sin }^{4}}\theta \left[ \frac{{{r}^{2}}}{\Sigma }-\frac{\Sigma -2{{r}^{2}}}{{{\Sigma }^{2}}}\left( {{r}^{2}}+{{a}^{2}}+\frac{{{a}^{2}}Mr{{\sin }^{2}}\theta }{\Sigma } \right) \right]{{\Omega }^{3}}\]
\[+ \left\{ M\frac{\Sigma -2{{r}^{2}}}{{{\Sigma }^{2}}}\left[ \frac{6Mr\left( {{r}^{2}}+{{a}^{2}} \right)}{\Sigma }+3\Delta -\Sigma  \right]+\left( 1-\frac{2Mr}{\Sigma } \right)r \right\} 
\]
\[\cdot \ {{\Omega }^{2}} {{\sin }^{2}}\theta-\frac{\Sigma -2{{r}^{2}}}{{{\Sigma }^{2}}}\frac{6a{{M}^{2}}r{{\sin }^{2}}\theta }{\Sigma }\Omega
\]
\begin{equation}
    +\Delta {{\sin }^{2}}\theta \; \Omega \frac{\partial \Omega }{\partial r}-M\frac{\Sigma -2{{r}^{2}}}{{{\Sigma }^{2}}}\left( 1-\frac{2Mr}{\Sigma } \right)=0 . 
    \label{Q}
\end{equation}
In order to solve Eq. (\ref{Q}), we must define the angular velocity function. We use here the angular velocity profile proposed and explained in \citet{FW04,FW07} and \citet{YWF}:
\begin{equation}
    \Omega \left( \varpi  \right)=\Omega \left( r \sin \theta  \right)=\frac{\sqrt{M}}{{{\left( r \sin \theta  \right)}^{{3}/{2}\;}}+a\sqrt{M}}{{\left( \frac{{{r}_{K}}}{r \sin \theta } \right)}^{{{n}_{p}}}} ,
    \label{U}
\end{equation}
where ${{r}_{K}}$ is the equatorial plane radius at which the material moves with Keplerian velocity. Here, the parameter ${{n}_{p}}$ corresponds to pressure forces and is responsible for the geometry of the torus determining its thickness. Tori solutions for various ${{r}_{K}}$ and ${{n}_{p}}$ values are shown in Fig. \ref{orsts}, while the selected tori used in our simulations are displayed in Fig. \ref{opaqdk}.\label{orstf}
\end{enumerate}

The last thing remaining for ADs of this kind is to calculate the frequency integrated specific intensity. This is done by using the method described in Sect. \ref{2.6} and then applying Sect. \ref{2.4} to determine the radiation flux or force and the ensuing acceleration caused by the disk's hot material.
    
   \begin{figure*}[t]
   \centering
   \includegraphics[width=\hsize]{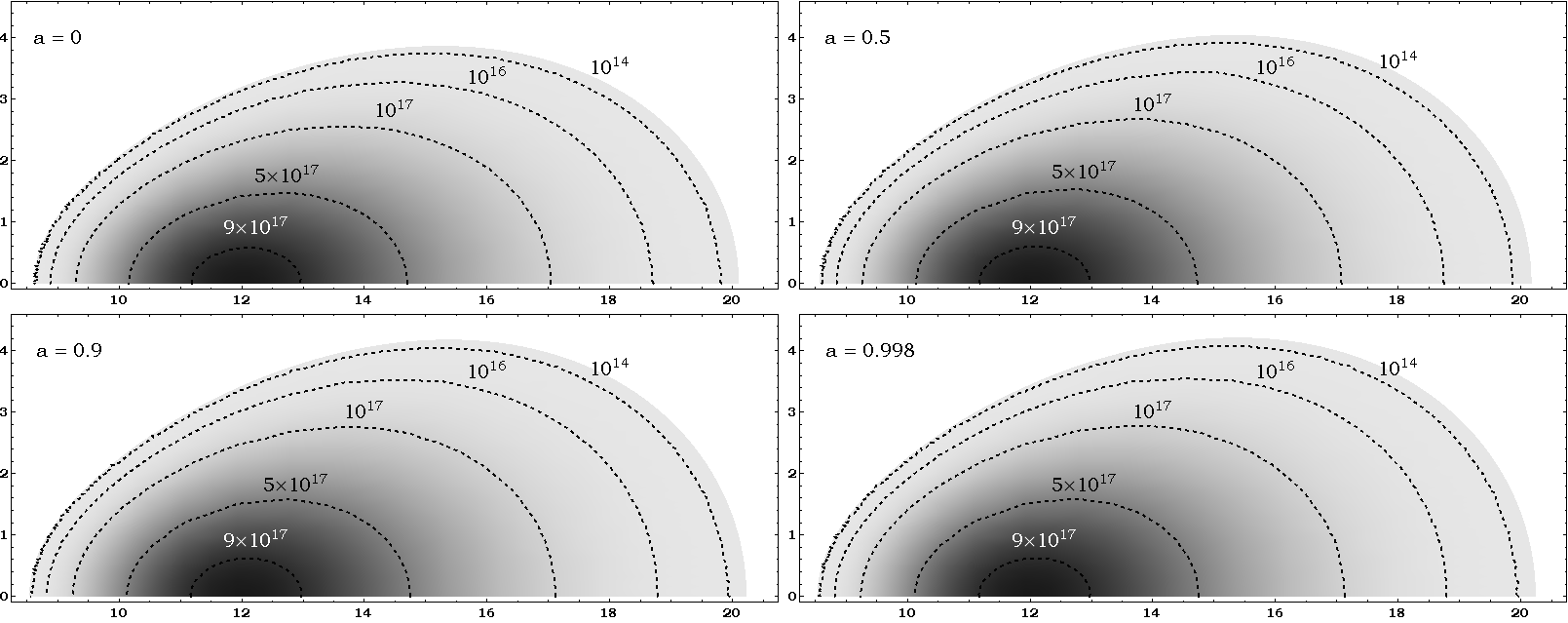}
      \caption{Polish doughnut number density cross sections (Sect. \ref{3.2}b, c). The center of all tori lies at $\left( 12, 0 \right)$ and has a number density of ${{10}^{18}} {\rm {cm}^{-3}}$. The top row shows the disk for BH spin parameters $a=0$ and $a=0.5$, while the bottom row for $a=0.9$ and $a=0.998$.
      }
         \label{pd_density_s}
   \end{figure*}
   
\subsection{Semi-opaque tori}
\label{3.2}
In this subsection we refer to the semi-opaque and transparent disk models considered in our work. Again as before, some of the models are more simplistic than others (toy or snapshot models) and some are based on specific physical conditions and are more complex (self-consistent models).

\begin{figure}[t]
   \centering
   \includegraphics[width=\hsize]{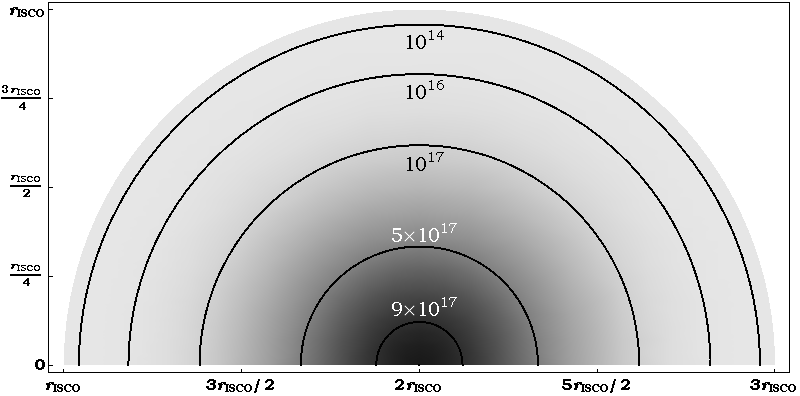}
      \caption{LFM model number density cross section for any spin parameter (Sect. \ref{3.2}d, e). The center of the tori lies at $\left( 2 {{r}_{\rm ISCO}}, 0 \right)$ and has a number density of ${{10}^{18}} {\rm {cm}^{-3}}$.
      }
         \label{LFM_num_den}
   \end{figure}

For our research, we considered up to this point five disk models that could fit in the semi-opaque or transparent category. Specified by the name of the considered models, we have the following cases:
\begin{enumerate}[label=(\alph*)]
\item No torus: there is no AD around the central BH. Photon trajectories continue until they either cross the event horizon or escape the system by crossing an adjustable outer radius boundary. We use this mode when we wish to study just geodesics for example.
\item Semi-opaque pressure supported polish doughnut (PS PD -- self-consistent model): a stationary and axisymmetric radiation pressure supported polish doughnut. We construct this accretion torus following \citealt{A78} and \citealt{K78}. As in previous tori, we assume here that the material has no significant radial or poloidal velocity components. The material thus has four-velocity ${{u}^{\alpha }}=\left( {{u}^{t}}, {{u}^{\phi }}, 0, 0 \right)$. We follow \citealt{YWF} and assume that the torus has a polytropic equation of state $P=\kappa n^{\Gamma}$, where $n$ the material number density,\break $\Gamma = 4/3$ and $\kappa =\hbar c{{\left\{ {45\left( 1-\beta  \right)}/{{{\left[ {{\pi }^{2}}\left( \mu  {{m}_{p}} \beta  \right) \right]}^{4}}}\; \right\}}^{{1}/{3}\;}}$, with $\hbar $ the reduced Planck constant, $\mu $ the mean molecular weight and $\beta $ the ratio of gas pressure to total pressure. We also assume that the disk rotation follows (\ref{U}). The AD is then described by:
\[{{\partial }_{r}}\xi \left( r, \theta  \right)=-{{a}_{r}}\left( r, \theta  \right),
\]
\begin{equation}
    {{\partial }_{\theta }}\xi \left( r, \theta  \right)=-{{a}_{\theta }}\left( r, \theta  \right) ,
\end{equation}
where $\xi \left( r, \theta  \right)=\ln \left[ \Gamma -1+\Gamma \kappa  n{{\left( r, \theta  \right)}^{\Gamma -1}} \right]$ is a function of the disk number density $n\left( r, \theta  \right)$ and the acceleration components are ${{a}_{\alpha }}\left( r, \theta  \right)={{u}_{\alpha ;\,\beta }} \left( r, \theta  \right)u{{\left( r, \theta  \right)}^{\beta }}$. The maximum number density point lies in the equatorial plane at ${{r}_{\rm center}}={{r}_{\rm K}}$ and the number density there is ${{n}_{\rm center}}={{10}^{18}}\: {{\rm cm}^{-3}}$. Cross sections of the tori number density used are shown in Fig. \ref{pd_density_s}.\label{pdb}
\item Translucent PS PD: a translucent pressure supported polish doughnut. It is the same as the above torus, but displays no absorption of photons by the disk material.\label{pdc}
\item Semi-opaque LFM torus (toy, snapshot model): a stationary and axisymmetric semi-opaque torus of circular cross section. The cross section center lies on the equatorial plane at ${{r}_{\rm center}}=2 {{r}_{\rm ISCO}}$ and the torus cross section has a radius ${{r}_{\rm torus}}={{r}_{\rm ISCO}}$. The torus thus stretches from an inner radius of ${{r}_{\rm inner}}={{r}_{\rm ISCO}}$ to an outer radius ${{r}_{\rm outer}}=3 {{r}_{\rm ISCO}}$ and has a maximum height ${{h}_{\rm torus}}={{r}_{\rm torus}}={{r}_{\rm ISCO}}$. The center number density is ${{n}_{\rm center}}={{10}^{18}} {{\rm cm}^{-3}}$ and decreases to zero moving toward the torus surface. For this torus the product ${{a}_{\nu }}\cdot 2 {{r}_{\rm torus}}$ is $\sim 1-5$ throughout the cross section. Images of the cross section number density are shown in Fig. \ref{LFM_num_den}.\label{lfmd}
\item Translucent LFM torus (toy, snapshot model): a translucent LFM torus of circular cross section. It is the same as the previous model, but without its material absorbing any of the photons crossing it.
\end{enumerate}

The disk models we discussed above are responsible for giving us most importantly the material's number density $n\left( r, \theta  \right)$. From that, we can then obtain other useful quantities for the matter, one of which is the material temperature. Following standard procedures, we have here:
\begin{equation}
    T\left( r, \theta  \right)=\frac{\hbar c}{k}{{\left[ \frac{45\left( 1-\beta  \right)}{{{\pi }^{2}}\mu  {{m}_{p}} \beta } \right]}^{{1}/{3}}}\rho {{\left( r, \theta  \right)}^{{1}/{3}}} ,
\end{equation}
where $\rho$ the (volumetric mass) density.

Continuing on, we can obtain firstly the necessary material's absorption coefficient from Eq. (\ref{a_nu R}) as:
\begin{equation}
    {{a}_{\nu }}\left( r, \theta  \right)={{\sigma }_{\nu }}\: n\left( r, \theta  \right) ,
\end{equation}
where ${{\sigma }_{\nu}}$ is the absorption cross section best chosen for the processes under study. Further on, we find the emission coefficient of the material by using the thermal emission and blackbody radiation properties we have for the assumed disk. The thermal emission assumption dictates that the emission coefficient will be given by:
\begin{equation}
    {{j}_{\nu }}\left( r, \theta  \right)={{a}_{\nu }}\left( r, \theta  \right) {{B}_{\nu }}\left( T \right) ,
\end{equation}
where ${{B}_{\nu }}\left( T \right)$ the Planck function and $T$ the corresponding temperature. In order to procure the Planck function, we make use of the blackbody attributes of the material and have:
\begin{equation}
    {{B}_{\nu }}\left( T \right)=\frac{{2h {{\nu }^{3}}}/{{{c}^{2}}}}{\exp \left( {h\nu }/{kT} \right)-1} ,
\end{equation}
where $k$ the Boltzmann constant. Combining then the above equations, we have for $j_{\nu}$ that:
\begin{equation}
    {{j}_{\nu }}\left( r, \theta  \right)={{\sigma }_{\nu }}\; n\left( r, \theta  \right)\frac{{2h {{\nu }^{3}}}/{{{c}^{2}}}}{\exp \left[ {h\nu }/{kT}\left( r, \theta  \right) \right]-1} .
\end{equation}
We additionally remark here that in the above equations, we could also add shaping functions to modify the emission and the absorption of the material. We could this way study other disk models or perhaps different physical properties. A more general form of the above functions could thus be:
\[{{a}_{\nu }}\left( r, \theta  \right)={{C}_{\rm abs}}\: {{\sigma }_{\nu }}\: f\left( n\left( r, \theta  \right), T\left[ n\left( r, \theta  \right) \right], E\left[ n\left( r, \theta  \right) \right] \right) ,\]
\begin{equation}
    {{j}_{\nu }}\left( r, \theta  \right)={{C}_{\rm em}}\: f\left( n\left( r, \theta  \right), T\left[ n\left( r, \theta  \right) \right], E\left[ n\left( r, \theta  \right) \right] \right) ,
\end{equation}
where ${C}_{\rm abs}$ and ${C}_{\rm em}$ are absorption and emission coefficients respectively and $f$ a shaping function of the number density $n\left( r, \theta  \right)$, the material temperature $T\left[ n\left( r, \theta  \right) \right]$ and the photon energy function $E\left[ n\left( r, \theta  \right) \right]$. We can see one such case for example in \citealt{YWF}.

After concluding the calculations described above, we have the resulting radiation intensity $I\left( r, \theta , \tilde{a}, \tilde{b} \right)$. We can then apply the method of Sect. \ref{2.4}, obtaining the stress – energy tensor, the flux and the force of the radiation.

   \begin{figure}[t]
   \centering
   \includegraphics[width=\hsize]{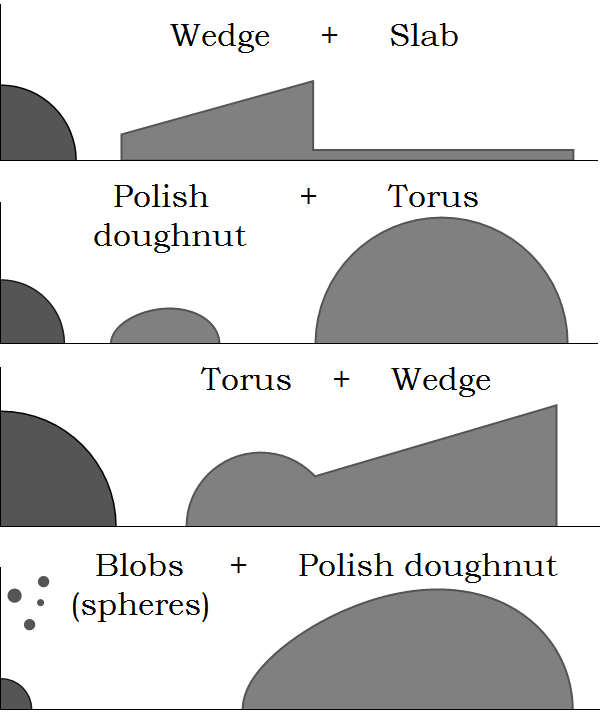}
      \caption{ADs created from combinations of previously referred models (\textit{from top to bottom}): wedge \& slab, polish doughnut \& torus, torus \& wedge and polish doughnut \& spheres.
      }
         \label{more_tori}
   \end{figure}

\section{Algorithms and codes}
\label{S4}
In this section we present the codes we developed from 2012 to 2021, used in our work and kept improved and cross checked since. We explain their capabilities and show some of their results. We also note here that all of the codes used and presented, were designed in order to be executed with extremely limited computational resources. This has important consequences on the design, speed and effectiveness required from the codes.

\subsection{Code \texttt{Omega}}
\label{Code Omega}
Code \verb+Omega+ was the first we created in our work and it is a central part of all following codes, as its main purpose is to calculate photon trajectories. It works for a Schwarzschild and a Kerr spacetime, but it can be easily modified in order to work in other spacetime models as well, for example Kerr – Newman, Reissner – Nordström, Friedmann – Lemaître – Robertson – Walker etc.

The code studies the tori models referred to and explained in Sect. \ref{3.1} and Sect. \ref{3.2} and depicted in the accompanying Figs. \ref{2D_disks} - \ref{LFM_num_den}. It is also possible to add new AD models to the code, such as disks resulting from combinations of the previously mentioned models. Examples of such tasks are shown in Fig. \ref{more_tori}. Additionally to the above, \verb+Omega+ has an option "Star", where instead of a noncentral accretion disk, it studies the radiation field produced by a central spherical star of isotropic and uniform radiation. We have mostly used this option so far in order to cross-check and validate the veracity of our results in comparison to other studies, as we see further on. This code feature can also be modified as well, in order to study objects more complex, such as nonspherical or nonuniformly emitting stars.
   \begin{figure}[t]
   \centering
   \includegraphics[width=\hsize]{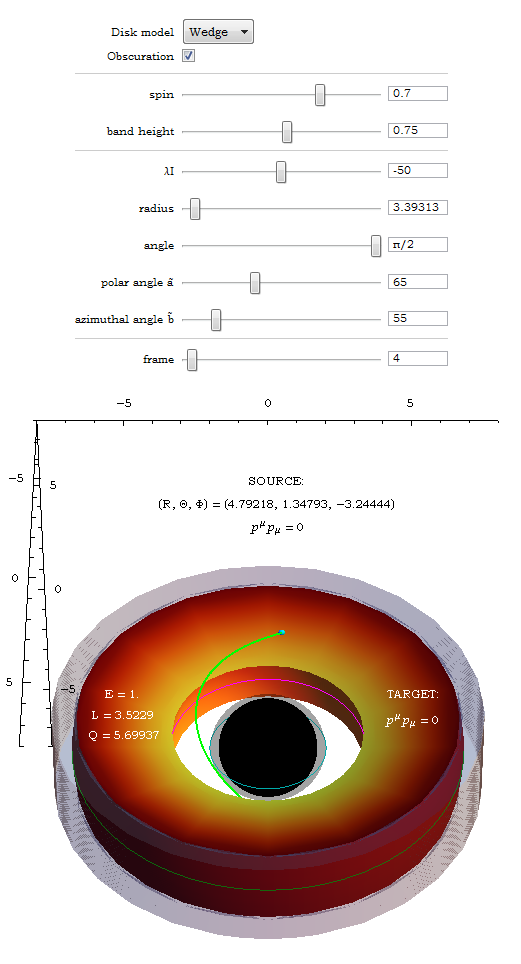}
      \caption{Control interface of the \texttt{Omega} code, as described in Sect. \ref{Code Omega}. In light green is the requested photon trajectory and with the cyan point we denote its point of origin.
      }
         \label{Omega}
   \end{figure}

\verb+Omega+ solves the particle trajectory equations mentioned in Sect. \ref{2.3} for a photon and finds the trajectory and the point of origin of said photon. This point of origin could be on the hot AD, the BH event horizon, or a location outside and far away from the system. If all that the backward photon trajectory intersects with is the event horizon or the system exterior, then no radiation or energy is carried to the AD material and the target particle.

   \begin{figure}[t]
   \centering
   \includegraphics[width=\hsize]{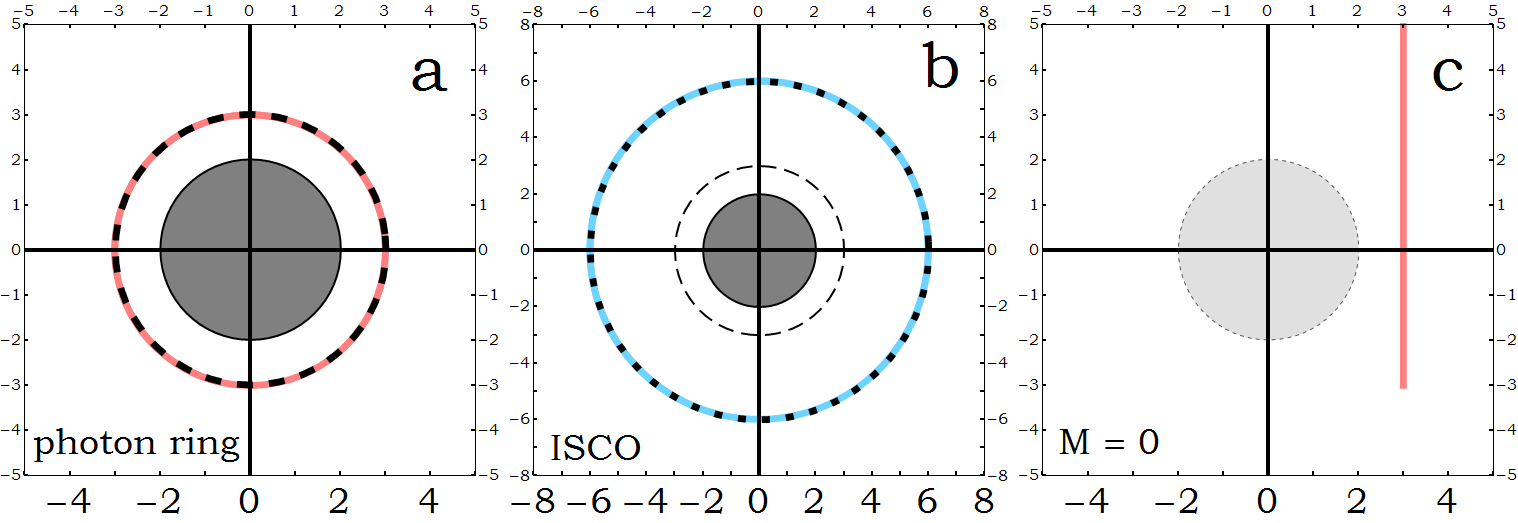}
      \caption{Tests for the validity of \texttt{Omega}. We confirm that the code successfully produces fundamental particle orbits, such as the ones mentioned in Sect. \ref{Kerr metric}. In each picture we have the BH situated at the axes origin and with gray is its event horizon. In (\textit{a}) we have in red the photon ring produced numerically by the code and on top of it, in the black dashed circle, its theoretically expected position. In (\textit{b}) we see in blue the ISCO produced numerically and on top of it, in the black dotted circle, its theoretically expected position. In (\textit{c}) we set the BH mass $M=0$ (with light gray showing where its horizon would have been) and examine the trajectory of a photon. With any BH mass $M>0$, the photon would have to infall and cross the event horizon, but with no BH mass present, the spacetime is Minkowski and the photon travels in a straight line.
      }
         \label{test trajectories}
   \end{figure}

   \begin{figure}[t]
   \centering
   \includegraphics[width=\hsize]{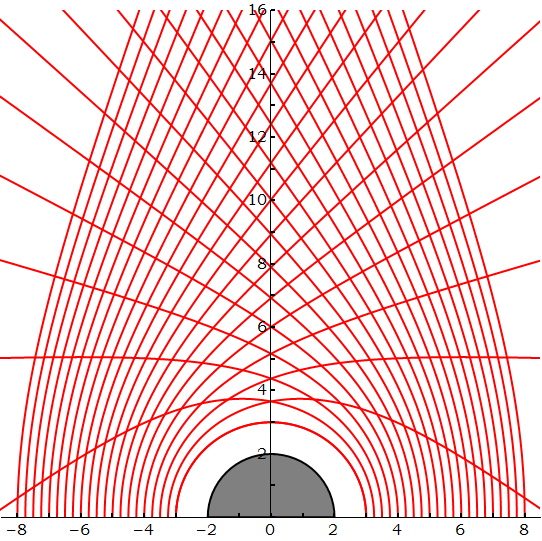}
      \caption{Free photon trajectories in the poloidal $\rm{x-z}$ plane. Photons are emitted perpendicularly upward from various points of the equatorial plane of the system. The closest to the BH emission points are $(3,0)$ and $(-3,0)$ and taking into account the angle of emission, these photons should and indeed do follow the photon ring trajectory \citep[compare to][Fig. 1]{Bini15}.
      }
         \label{m2b}
   \end{figure}

Depending on the environment in which we use it, \verb+Omega+ can have different outputs. In its original form, which is visual, the program has an interface that allows the user to select primary properties for the environment, such as the disk model, the BH spin parameter and the disk height. Also, the user can select important options for a trajectory, including its maximum length, its point of origin and angle of emission and the two emission angles $\tilde{a}$ and $\tilde{b}$. Finally, there are some additional visualization options that include the choice of frame size of the visual box and the depiction of obscured parts of the outer disk. The code's dynamic output picture shows the BH event horizon and its ergosphere, the AD and the requested photon trajectory. The particle trajectory is drawn in different styles and colors for escaping photons, photons infalling in the BH and photon trajectories starting from the AD. In addition, some trajectory information are displayed in the picture, including the photon energy and angular momentum, the trajectory's Carter constant and a confirmation of the photon momentum magnitude conservation. All the above can be seen in Fig. \ref{Omega}.
   
The latest \verb+Omega+ versions are the most compact, since they now are functions and parts of more complex codes, outputting thus only numerical data and no visual information. \verb+Omega+ runs for a single photon trajectory per execution and returns key information for this trajectory. Firstly, it reveals the existence or absence of incoming radiation in the requested direction. Additionally, it gives the precise photon trajectory numerically as well as the coordinates of the emission source.

In order to verify the validity of the calculated particle trajectories, we subjected \verb+Omega+ to various tests. Firstly, we verified that \verb+Omega+ can successfully predict elementary yet fundamental particle trajectories, such as the notable orbits presented in \citet{B72} and discussed in Sect. \ref{Kerr metric}. In Fig. \ref{test trajectories} we see that we indeed obtain such orbits, like the photon ring (massless particles) and the ISCO (massive particle). Additionally, we confirm that if the BH mass is reduced to zero, the spacetime is no longer a curved Schwarzschild or Kerr one. Instead, it degenerates into a flat Minkowskian form, causing the particle orbits to turn into straight lines.

Continuing on, we compare our code results to the work of \citealt{Bini15} concerning photon trajectories and we note agreement. In Fig. \ref{m2b} we see for example free photon trajectories produced by \verb+Omega+. The photons are launched perpendicularly and upward from various points in the equatorial plane of a Schwarzschild BH \citep[we compare to][Fig. 1]{Bini15}.

\begin{figure}[t]
   \centering
   \includegraphics[width=\hsize]{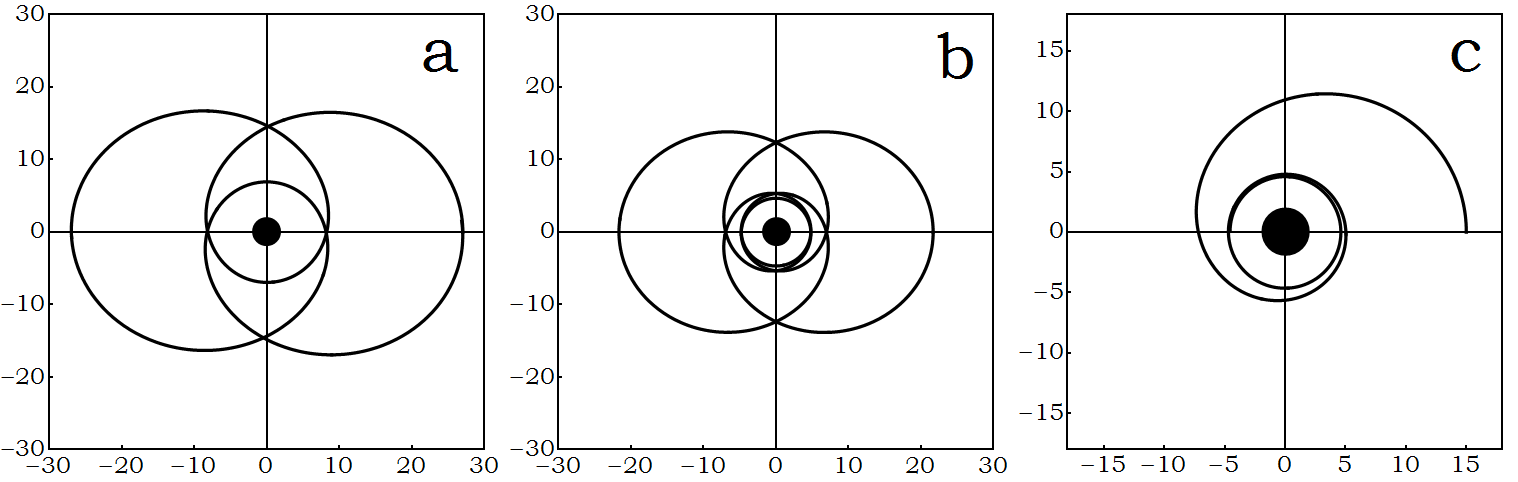}
      \caption{\texttt{Omega} code successfully creating massive particle orbits. These trajectories are in agreement with the corresponding orbits shown in \citet{LPG}.}
         \label{Levin pics}
   \end{figure}

Finally, we examine massive particle trajectories and compare \verb+Omega+ results to the work of \citet{LPG} and \citet{LG}, with which we are also in accord. In Fig. \ref{Levin pics}, for example, we cross-check our code results with what is shown in \citet{LPG} and we again notice agreement.

\subsection{Code \texttt{Infinity}}
\label{Code Infinity}

   \begin{figure}[!t]
   \centering
   \includegraphics[height=0.74
   \textheight,keepaspectratio]{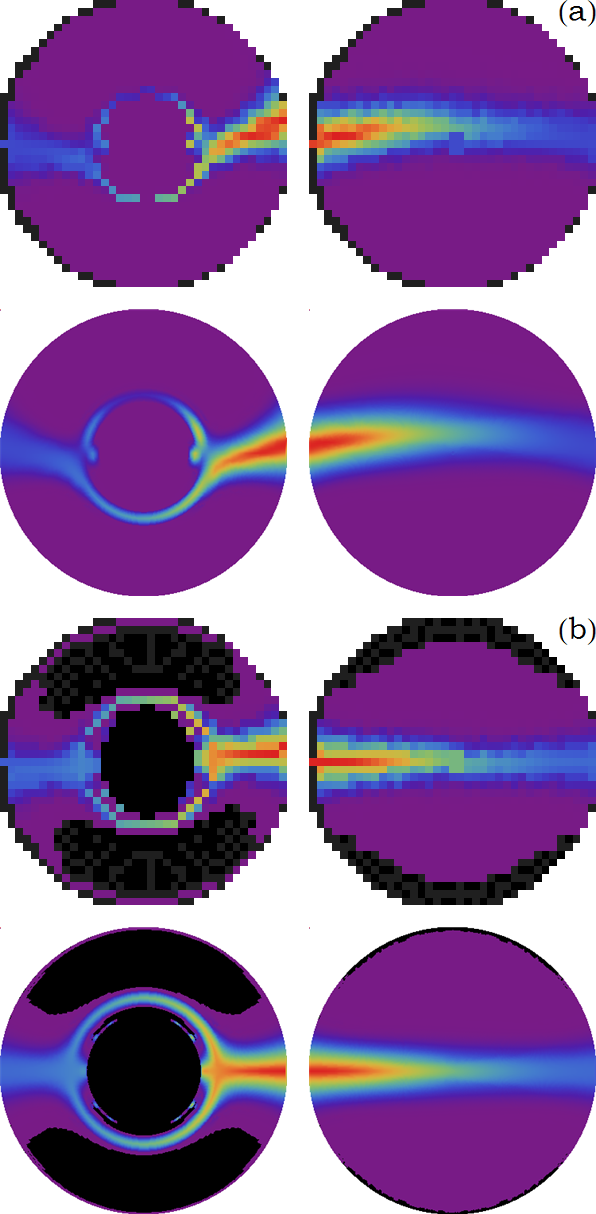}
      \caption{Two instances of the first version of the resolution enhancement process ($625$ times here). In the \textit{top row} of each part is the original image created by the AD observation. On the \textit{bottom} is the resulting image after the resolution enhancement. These functions are now rewritten and incorporated into \texttt{Infinity} to enhance results by a number of times.
      }
         \label{magic}
   \end{figure}

The \verb+Infinity+ code is now an extended and fully automatic program that can run complete simulations for a large stream of different points and situations. These runs do not require to be for the same object or of the same environment. It automatically creates output files and saves numerical data, images and information possibly useful for other runs or aggregated results. It also has features to prevent the loss of data and computational time during power or network outages.

   \begin{figure}
   \centering
   \includegraphics[width=\hsize]{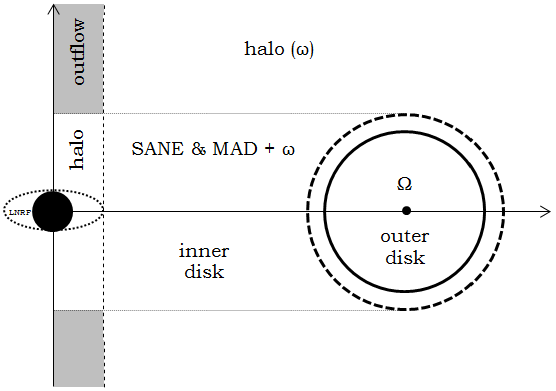}
      \caption{Regions of a system we consider in a run. “Halo” only includes observers at rest in the local rest frame and rotating with $\omega$. “Outer disk” is the main AD of the system. Material there can move in four different ways: at rest in the local rest frame rotating with $\omega$, purely azimuthally with angular velocity $\Omega$, or by slowly infalling in a manner like SANE and MAD. “Inner disk” is the region of infalling material and thus includes only matter at rest in the local rest frame and infalling material mimicking SANE and MAD. The “Outflow” region is part of a cylinder centered at the system's rotation axis that includes the BH's ergosphere and has thus a radius $r=2M$. The actual region is considered to exist above a certain height, further away from the BH. In this region, we can have matter at rest in the local rest frame and matter flowing outward with a certain velocity. We may also consider a subregion there (see e.g., \citealt{Asada16,Park19}), a narrower cone or cylinder with radius $r=M$ with stronger outflow and faster velocity or even with velocity of the opposite direction.
      }
         \label{disk_velocities}
   \end{figure}

\begin{table}
\centering \caption{Comparison of \texttt{Infinity} results with formulae in Schwarzschild spacetime}

\begin{tabular}{ccccc}
 \hline
 \hline
 \text{Star} & \text{Orbit} & \text{Approx.} & \text{Formula} & ${{T}^{\hat{\alpha}\hat{\beta}}}$ \\
 \text{radius} & \text{radius} & \text{radius} & \text{radius} & \text{errors} \\
 \text{R (M)} & \text{r (M)} & \text{(${}^{\circ}$)} & \text{(${}^{\circ}$)} & \text{($\%$)} \\
 \hline
 4 & 4.1 & 80.8 & 80.9 & 0.17 \\
 4 & 5 & 61.2 & 61.2 & 0.65 \\
 4 & 6 & 50. & 50.3 & 0.62 \\
 4 & 7 & 42.8 & 43.1 & 0.44 \\
 4 & 8 & 37.6 & 37.8 & 0.25 \\
 5 & 6 & 61.2 & 61.5 & 0.17 \\
 5 & 8 & 44. & 44.3 & 0.7 \\
 5 & 10 & 35.2 & 35.3 & 0.98 \\
 6 & 7 & 62.4 & 62.5 & 0.23 \\
 6 & 8 & 52.4 & 52.7 & 0.43 \\
 6 & 9 & 46. & 46.1 & 0.71 \\
 6 & 10 & 40.8 & 41.1 & 0.55 \\
 \hline
\end{tabular}

\tablefoot{
Radiation from a nonrotating central star of radius $R$ in Schwarzschild spacetime as recorded by an observer in orbit of radius $r$. The approximate optical radius in degrees (${}^{\circ}$) of the star given by \texttt{Infinity} is in satisfying agreement with the star optical radius given by the formula in \citet{AEL90}. We also see the errors percent of the code calculations of the stress--energy tensor compared to the relevant formula of the aforementioned work.
}

\label{AELtab}
\end{table}

In the \verb+Infinity+ code, we choose a point of interest anywhere in the system under study and the program sets an observer there. It then scans the entire sky around this observer, solving the GRRTE and returning the radiation reaching the observer from each of the angles of the local sky. It then calculates the radiation stress – energy tensor and creates various images and sky maps of the radiation. It continues on to run a procedure that increases the code resolution by a significant number of times (see Fig. \ref{magic}).

   \begin{figure}
   \centering
   \includegraphics[width=\hsize]{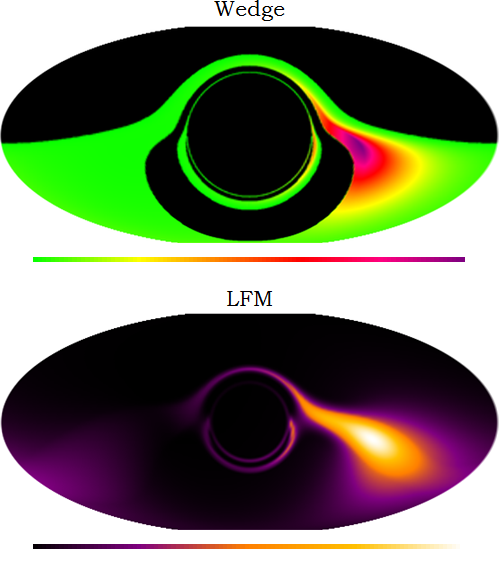}
      \caption{\texttt{Infinity} code: Mollweide sky maps of the frequency integrated specific intensity for tori with downward angular velocity vector. The color scale for each image is displayed below it, from minimum to maximum. \textit{Top}: Wedge AD model (opaque, Sect. 3.1d), \textit{bottom}: LFM AD model (semi-opaque, Sect. 3.2d).
      }
         \label{mollweides short}
   \end{figure}

The program continues on to calculate the radiation flux and the force four-vectors applied to various observers. This includes observers at rest in the local frame rotating with ${\omega =\omega \left( r, \theta  \right)}$ (Eq. \ref{S}) as seen from infinity, observers in circular orbits moving with $\Omega $ (Eqs. \ref{T} and \ref{U}) depending on the selected disk model and observers on accreting material approximating the SANE and MAD models. In addition to the above, we can also have observers in two possible outflow regions, "jet regions" if you will, close to the system rotation axis and above a certain height (Fig. \ref{disk_velocities}). Let us note here that, even though we have made our choice on certain velocity profiles for the various disk models, this can easily be adjusted to suit the needs of other disk models with different material velocities.
   
The program then concludes by outputting its results and creating various save files. These include a Mollweide map picture (Fig. \ref{mollweides short}) for the radiation in the specific observer location, pictures of the important matrices of the simulation and tables of the assorted profiles' force components and the radiation stress – energy tensor matrix. Additionally to the figures, videos of flights around and if possible through the disk for all the models we studied, can be found on youtube.com, under the name of this work's creator "Leela Elpida Koutsantoniou"\footnote{Direct link: \url{youtube.com/channel/UCJ4v8rSg390gt9kQfVtfx5A}}.

\begin{table}[!ht]
\centering \caption{Comparing \texttt{Infinity} with formulae in Kerr spacetime}

\begin{tabular}{cccc}
 \hline
 \hline
 \text{Star} & \text{Orbit} & \text{fw radius} & \text{bk radius} \\
 \text{radius} & \text{radius} & \text{error} & \text{error} \\
 \text{R (M)} & \text{r (M)} & \text{($\%$)} & \text{($\%$)} \\
 \hline
 \multicolumn{4}{c}{$a=0.2$}\\
 4 & 4.1 & 0. & 0.1 \\
 4 & 6 & 0.4 & 0.2 \\
 4 & 7 & 0. & 0.4 \\
 4 & 8 & 0. & 0.3 \\
 4 & 9 & 0. & 0.3 \\
 5 & 5.1 & 0.1 & 0.1 \\
 5 & 6 & 0.2 & 0. \\
 5 & 9 & 0. & 0. \\
 5 & 10 & 0.3 & 0. \\
 6 & 6.1 & 0.1 & 0. \\
 6 & 8 & 0.2 & 0.2 \\
 6 & 10 & 0.5 & 0. \\
 \hline
 \multicolumn{4}{c}{$a=0.5$}\\
 4 & 4.1 & 0.4 & 0.1 \\
 4 & 6 & 0.4 & 0.5 \\
 5 & 6 & 0. & 0.3 \\
 5 & 7 & 0.6 & 0. \\
 5 & 8 & 0.5 & 0.4 \\
 5 & 9 & 0.3 & 0.2 \\
 5 & 10 & 0.3 & 0.3 \\
 6 & 6.1 & 0. & 0. \\
 6 & 7 & 0. & 0.2 \\
 6 & 8 & 0. & 0.4 \\
 6 & 9 & 0.2 & 0.2 \\
 6 & 10 & 0.5 & 0. \\
 \hline
 \multicolumn{4}{c}{$a=0.7$}\\
 4 & 4.1 & 0.3 & 0.2 \\
 5 & 5.1 & 0.4 & 0.1 \\
 5 & 6 & 0.7 & 0.6 \\
 5 & 7 & 0.8 & 0.4 \\
 5 & 8 & 0.7 & 1.1 \\
 5 & 9 & 0.8 & 0.7 \\
 5 & 10 & 1.5 & 0.3 \\
 6 & 6.1 & 0.2 & 0.1 \\
 6 & 7 & 0.3 & 0.5 \\
 6 & 8 & 0.2 & 0. \\
 6 & 9 & 0.7 & 0.6 \\
 6 & 10 & 0.8 & 0.5 \\
 \hline
 \multicolumn{4}{c}{$a=0.9$}\\
 4 & 4.1 & 0.6 & 0.7 \\
 5 & 5.1 & 0.3 & 0. \\
 5 & 6 & 1.4 & 0.8 \\
 5 & 7 & 1.7 & 1.4 \\
 5 & 8 & 1.7 & 1.7 \\
 5 & 9 & 2. & 1.6 \\
 5 & 10 & 2.5 & 1.6 \\
 6 & 6.1 & 0.5 & 0.4 \\
 6 & 7 & 0.5 & 0.8 \\
 6 & 8 & 0.6 & 0.2 \\
 6 & 9 & 0.9 & 1. \\
 6 & 10 & 1.3 & 1.1 \\
 \hline
\end{tabular}

\tablefoot{Examining the concurrence between \texttt{Infinity} results and the formulae given in \citet{LM95} and \citet{ML96} for the source's forward (fw) and backward (bk) optical radius.}

\label{LMtab}
\end{table}

\begin{table}
\centering \caption{\texttt{Infinity} code convergence with resolution}

\begin{tabular}{cccccc}
 \hline
 \hline
 \text{Star} & \text{Orbit} & \text{Reso-} & \text{Approx.} & \text{Formula} & \text{Calcul.} \\
 \text{radius} & \text{radius} & \text{lution} & \text{radius} & \text{radius} & \text{error} \\
 \text{R (M)} & \text{r (M)} & \text{(${}^{\circ}$)} & \text{(${}^{\circ}$)} & \text{(${}^{\circ}$)} & \text{($\%$)} \\
 \hline
 4 & 5 & 10 & 60 & 61.2 & 2. \\
 4 & 5 & 5 & 60 & 61.2 & 2. \\
 4 & 5 & 2 & 60 & 61.2 & 2. \\
 4 & 5 & 1 & 61 & 61.2 & 0.3 \\
 4 & 5 & 0.4 & 61.2 & 61.2 & 0. \\
 \hline
 4 & 8 & 10 & 30 & 37.8 & 20.6 \\
 4 & 8 & 5 & 35 & 37.8 & 7.3 \\
 4 & 8 & 2 & 36 & 37.8 & 4.7 \\
 4 & 8 & 1 & 37 & 37.8 & 2. \\
 4 & 8 & 0.4 & 37.6 & 37.8 & 0.4 \\
 \hline
 4 & 9 & 10 & 30 & 33.7 & 10.9 \\
 4 & 9 & 5 & 30 & 33.7 & 10.9 \\
 4 & 9 & 2 & 32 & 33.7 & 4.9 \\
 4 & 9 & 1 & 33 & 33.7 & 2. \\
 4 & 9 & 0.4 & 33.6 & 33.7 & 0.2 \\
 \hline
 5 & 6 & 10 & 60 & 61.5 & 2.4 \\
 5 & 6 & 5 & 60 & 61.5 & 2.4 \\
 5 & 6 & 2 & 60 & 61.5 & 2.4 \\
 5 & 6 & 1 & 61 & 61.5 & 0.7 \\
 5 & 6 & 0.4 & 61.2 & 61.5 & 0.4 \\
 \hline
 5 & 9 & 10 & 30 & 39.2 & 23.5 \\
 5 & 9 & 5 & 35 & 39.2 & 10.8 \\
 5 & 9 & 2 & 38 & 39.2 & 3.2 \\
 5 & 9 & 1 & 39 & 39.2 & 0.6 \\
 5 & 9 & 0.4 & 39.2 & 39.2 & 0.1 \\
 \hline
 5 & 10 & 10 & 30 & 35.3 & 14.9 \\
 5 & 10 & 5 & 35 & 35.3 & 0.7 \\
 5 & 10 & 2 & 34 & 35.3 & 3.6 \\
 5 & 10 & 1 & 35 & 35.3 & 0.7 \\
 5 & 10 & 0.4 & 35.2 & 35.3 & 0.2 \\
 \hline
 6 & 7 & 10 & 60 & 62.5 & 4. \\
 6 & 7 & 5 & 60 & 62.5 & 4. \\
 6 & 7 & 2 & 62 & 62.5 & 0.8 \\
 6 & 7 & 1 & 62 & 62.5 & 0.8 \\
 6 & 7 & 0.4 & 62.4 & 62.5 & 0.2 \\
 \hline
 6 & 9 & 10 & 40 & 46.1 & 13.2 \\
 6 & 9 & 5 & 45 & 46.1 & 2.3 \\
 6 & 9 & 2 & 46 & 46.1 & 0.1 \\
 6 & 9 & 1 & 46 & 46.1 & 0.1 \\
 6 & 9 & 0.4 & 46. & 46.1 & 0.1\\
 \hline
\end{tabular}

\tablefoot{The decrease of calculation errors of central object optical radius by \texttt{Infinity} for increasing observation resolution in degrees (${}^{\circ}$).}

\label{incrres}
\end{table}

In order to certify the legitimacy of the code, we cross-checked our results with other works. We start by assuming that the radiation source is a central nonrotating star that emits photons isotropically from each point of its surface. We then assume that the star can have various radii and set observers at assorted distances from its surface. We then use \verb+Infinity+ and observe the system, noting the apparent size of the central object and the ensuing radiation stress--energy tensor. Finally, we look into our results and compare them to their expected values, given by analytical formulae in \citet{AEL90}. As we can see in Table \ref{AELtab}, we are in good agreement with the theoretically expected values. If, however, we desire so, we can have even better estimations by increasing the code resolution.

Keeping then in mind that in general the radiation source is not expected to be static but instead rotating, we expand the above check and look into environments around rotating stars. We thus look into rotating stars in Kerr spacetime in Table \ref{LMtab} and verify that \verb+Infinity+ does give good approximations of the central object dimensions, as given using formulae in \citet{LM95} and \citet{ML96}. Finally, in Table \ref{incrres}, we show the diminution of calculation errors with increasing code resolution. Even though a resolution of 0.4 degrees proves to be satisfactory to produce results of good quality, if the need arises for results of greater precision, \verb+Infinity+ can simply be executed with a higher resolution.

   \begin{figure*}
   \centering
   \includegraphics[width=\hsize]{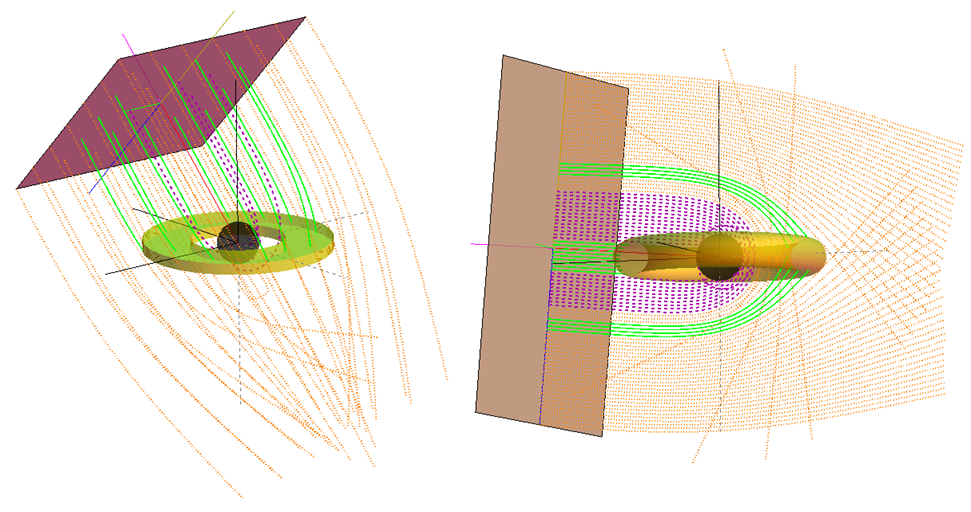}
      \caption{Images that show the workings of the \texttt{Elysium} code. On the \textit{left} is a run for a $7\times 7$ pixels screen with $j=0.5$, at an inclination $45{}^\circ $. On the \textit{right} is the center pixels column of a $100\times 100$ screen with $j=0$ at inclination $85{}^\circ $. The solid green lines are light rays that intersect the AD and thus carry radiation. The dashed purple lines end up in the event horizon and the dotted orange ones at infinity, both without crossing any part of the AD and thus attributing nothing to the radiation total.
      }
         \label{Elysium}
   \end{figure*}

\subsection{Code \texttt{Elysium}}
\begin{figure}
   \centering
   \includegraphics[width=\hsize]{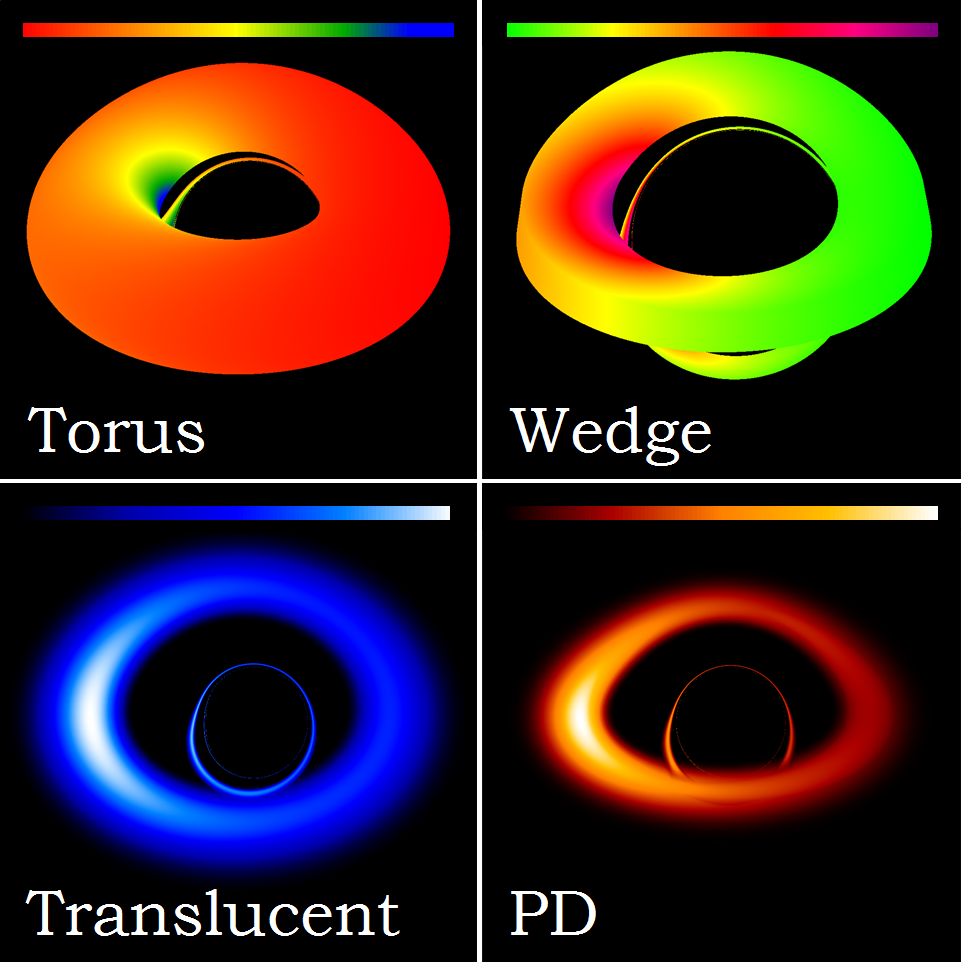}
      \caption{\texttt{Elysium} results: Pictures showing radiation specific intensity for four different AD models, as mentioned in Sect. \ref{S3}: torus (opaque, Sect. \ref{3.1}e), wedge (opaque, Sect. \ref{3.1}d), translucent (semi-opaque, Sect. \ref{3.2}c) and PD (semi-opaque, Sect. \ref{3.2}b). Radiation intensity color legends shown on the top from minimum to maximum. The second row pictures are similar to those shown in \citet{YWF}.
      }
      \label{2x2 pic}
   \end{figure}

\verb+Elysium+ code has as its main purpose to design and create a recording screen a specifically user-selected distance away from the BH and the AD system. Depending on the selected program resolution, the screen has the corresponding amount of “pixels” and from each of those, a light ray is emitted perpendicularly to the screen and moves toward the disk and the BH (Fig. \ref{Elysium}). Depending on what this ray will meet along its path, it returns information about its origin and the radiation received. We can see some examples in Fig. \ref{2x2 pic}. The pictures of the second row, as well as more that will be shown further on, agree with some of the assorted pictures shown in \citet{YWF}.
 
We remark here, that \verb+Elysium+ appears similar to the aforementioned \verb+Infinity+ code but is decidedly different, since it is practically its complement. \verb+Elysium+ therefore has equal quality and quantity of capabilities, options and result information as \verb+Infinity+. The difference of the two codes is literally, as well as figuratively, a point of view. \verb+Infinity+ starts at the single end point of ray trajectories and integrates the equations going backwards in time. This way, it finds if the intersecting light pathway can possibly traverse a light source at any point inside the system in question. \verb+Elysium+, on the other hand, has multiple starting points, the “pixels” of the screen, and integrates the equations of motion to see if an observer sitting at the pixel's location can see any part of the disk. There are various advantages and disadvantages in both methods and depending on the type of information required each time, we can choose the most appropriate and fast code of the two.

The reason why we deemed it worthy to present and discuss this imaging code is because this code turns out to be much more than a simple imaging mechanism. The information we can gather from its various processes and results can deliver important facts about the system under study. These include information about the central BH spin (see also Sect. \ref{Tranquillity}), the system mass and its distribution, the angular momentum allocation and flow and other such useful data. 

\verb+Elysium+ is the best choice to make when for example we want to study or compare to results of BH observations created using GRMHD simulations, or perhaps to construct the anticipated results these will give. One such example where we could utilize \verb+Elysium+ is to study bright, hot material orbiting in close proximity to the central BH, such as the cases described and discussed in \citealt{BLJun05,BLSep05}. We could also use it to construct the expected finite resolution disk images of an AD with various inclinations, similarly to what is presented in \citealt{Noble07}. Additionally, we could use it to investigate the radiative status of accretion flows, such as in \citealt{Moscibrodzka09}, the observational appearance of radiatively inefficient accretion flows (RIAFs) and jet outflows such as in \citealt{Moscibrodzka14}, or even the expected images of the SMBC in the Galactic center and its outflow regions, examined in \citealt{Moscibrodzka18}. Finally, we could use it to examine possible ways to test the legitimacy of the Kerr metric or even General Relativity in the vicinity of compact objects, such as described and examined in \citealt{BLJul06}.

\subsection{Code \texttt{Tranquillity}}
\label{Tranquillity}
The main purpose of \verb+Tranquillity+ is to find the AD inclination and then use it to make an assessment of the central BH spin. This is done by using information given by the incoming radiation emitted from the AD and traveling closer and further away from the BH and its gravitational well.

In the code's execution, an observer at infinity looks toward a BH and the AD swirling around it. The observer records in very high resolution and measures the relative position of the AD and its first Einstein ring\footnote{An Einstein, Chwolson or echo ring is created when light from the hot disk passes close to the BH and reaches an observer. Due to the BH-AD setup we study here, the relative locations of these objects causes the gravitational lensing taking place in the system to display a full ring of light, created by the accretion disk's upper and lower parts.}, the gravitationally lensed higher-order image, produced by photons circling around the BH. For this Einstein ring we henceforth use the designation "echo ring"\footnote{We refer to this ring as an echo due to its similarities with the light echo phenomenon, that is analogous to the sound echo, but with light instead of sound. The light echo feature, however, is more often connected to nova and supernova events (e.g., V838 Monocerotis, 2002 outburst), than with BHs (e.g., V404 Cygni, 2015 outburst).}, as this term is more intuitive, making the phenomenon easier to perceive and visualize. If the BH is nonrotating, the shapes of the AD and the echo will be concentric. If the BH is however rotating, there will be a divergence of the echo's center, tied to the BH spin. This is more or less expected since the photons forming the echo have to travel much closer to the BH and the perturbed spacetime around it. Therefore, the higher the BH spin is, the higher this divergence will be.

The first step of \verb+Tranquillity+ is to determine the inclination of the AD plane compared to the line of sight. Then, the code calculates the aforementioned echo divergence compared to the AD image. Finally, by using the inclination and divergence results, we can have an estimation of the central BH spin.

\begin{table}[!tp]
\centering \caption{\texttt{Tranquillity} code inclination estimation errors}
\begin{tabular}{llllll}

 \hline
 \hline
\text{\;$\iota$   \textbackslash \; j} & \text{0} & \text{0.2} & \text{0.5} & \text{0.7} & \text{0.9} \\
\text{\text{(${}^{\circ}$)}} & \text{(${}^{\circ}$)} & \text{\,(${}^{\circ}$)} & \text{\,(${}^{\circ}$)} & \text{\,(${}^{\circ}$)} & \text{\,(${}^{\circ}$)} \\
 \hline
$0$  &  0. &  0. &  0. &  0. &  0.\\ 
$10$ &  0.64 &  0.4 &  0.22 &  0.19 &  0.07\\ 
$20$ &  1.38 &  0.25 &  0.96 &  0.68 & 0.55\\
$30$ &  1.85 &  1.28 &  1.08 &  0.81 & 0.68\\
$40$ &  1.85 &  1.47 &  1.11 &  0.89 & 0.69\\
$50$ &  1.85 &  1.2 &  1.06 &  0.92 &  0.67\\
$60$ &  1.73 &  1.25 &  0.95 &  0.78 &  0.56\\
$70$ &  1.34 &  0.99 &  0.77 & 3.87 &  2.00\\
$80$ &  0.55 &  0.53 &  0.44 &  0.35 & 0.37\\
$90$ &  0. &  0. &  0. &  0. &  0.\\
 \hline

\end{tabular}

\tablefoot{Declinations of the \texttt{Tranquillity} code estimation of the disk inclinations for various BH spins (columns) and inclinations (lines). The spin is shown in the top row and the inclination, set by hand for the disk, is measured in degrees and is displayed in the first column. The values in the matrix are the errors in the \texttt{Tranquillity} code execution and are also measured in degrees. Two of the three worse cases of erroneous results, caused by the echo ring, are shown here, but are still rather small.}

\label{Table2}
\end{table}

The calculations for the AD inclination were rather successful, since the average declination was below 0.8 degrees for the 240 cases of different inclinations and spin parameters examined (see Table \ref{Table2}). Nonetheless, about 1.5\% of the cases examined, gave inclination errors above average. The estimation errors were about five degrees or less, and were caused by specific conditions of the setup that had a BH first echo ring appear in very particular and peculiar locations. The model selected for the AD of the object in question, does not appear to play any significant part in the inclination assessment so far.

An important note here is that the disk model adopted plays a very important part in the divergence calculations. Since the inner edge of the disk examined can be at very different radii, depending on the model examined, different disk models will follow slightly different divergence plots, directly affected by the disk's inner edge radius \citep[see][]{A10}.
We clearly state here however, that our code does not use or rely at all on the inner edge of the disk, but only on the appearance of the entire disk as a whole. The disk model thus, does not prove to be an insurmountable problem, since it only causes a small recalibration to the divergence plot, as we see in the second part of this work.

\subsection{Code \texttt{Burning Arrow}}
The \verb+Burning Arrow+ code has as a main purpose to study the BH massive particles orbit degradation due to the hot disk radiation.

In order to study the particle motion, the code must solve the general relativistic equations of motion, equivalent to the Classical Newton's Laws of Motion. Starting from the first law, in absence of general relativistic forces, the particle in question will follow a geodesic through spacetime. This geodesic obeys the equation:
\begin{equation}
    \frac{{{{\rm d}}^{2}}{{x}^{\mu }}}{{\rm d}{{\tau }^{2}}}+\Gamma _{\alpha \beta }^{\mu } {{u}^{\alpha }}{{u}^{\beta }}=0 ,
\end{equation}
where $\tau $ here is the proper time, ${{u}^{\alpha }}={\rm d}{{x}^{\alpha }}/{{\rm d}\tau }$ is the four-velocity and $\Gamma _{\alpha \beta }^{\mu }=\frac{1}{2}{{g}^{\mu \nu }}\left( {{g}_{\alpha \nu ,\beta }}+{{g}_{\beta \nu ,\alpha }}-{{g}_{\alpha \beta ,\nu }} \right)$ the Christoffel symbols, with commas used for partial derivatives. As in Newton's equation, the zero term on the right stands for the absence of acceleration. Solving the above, gives the various geodesic solutions that describe among others, special circular orbits such as the ISCO, the photon sphere etc. The previous equation can also be rewritten as:
   
\begin{equation}
    \frac{{{{\rm d}}^{2}}{{x}^{\mu }}}{{\rm d}{{t}^{2}}}+\Gamma _{\alpha \beta }^{\mu } \frac{{\rm d}{{x}^{\alpha }}}{{\rm d}t}\frac{{\rm d}{{x}^{\beta }}}{{\rm d}t}-\Gamma _{\alpha \beta }^{t}\frac{{\rm d}{{x}^{\alpha }}}{{\rm d}t}\frac{{\rm d}{{x}^{\beta }}}{{\rm d}t}\frac{{\rm d}{{x}^{\mu }}}{{\rm d}t}=0 ,
\end{equation}
by using the chain rule in order to have derivatives by the coordinate time ${{x}^{0}}=t$. This, nevertheless, has proved to be more prone to error accumulation in our work and so we choose to work with the proper time equations instead.

   \begin{figure}
   \centering
   \includegraphics[width=\hsize]{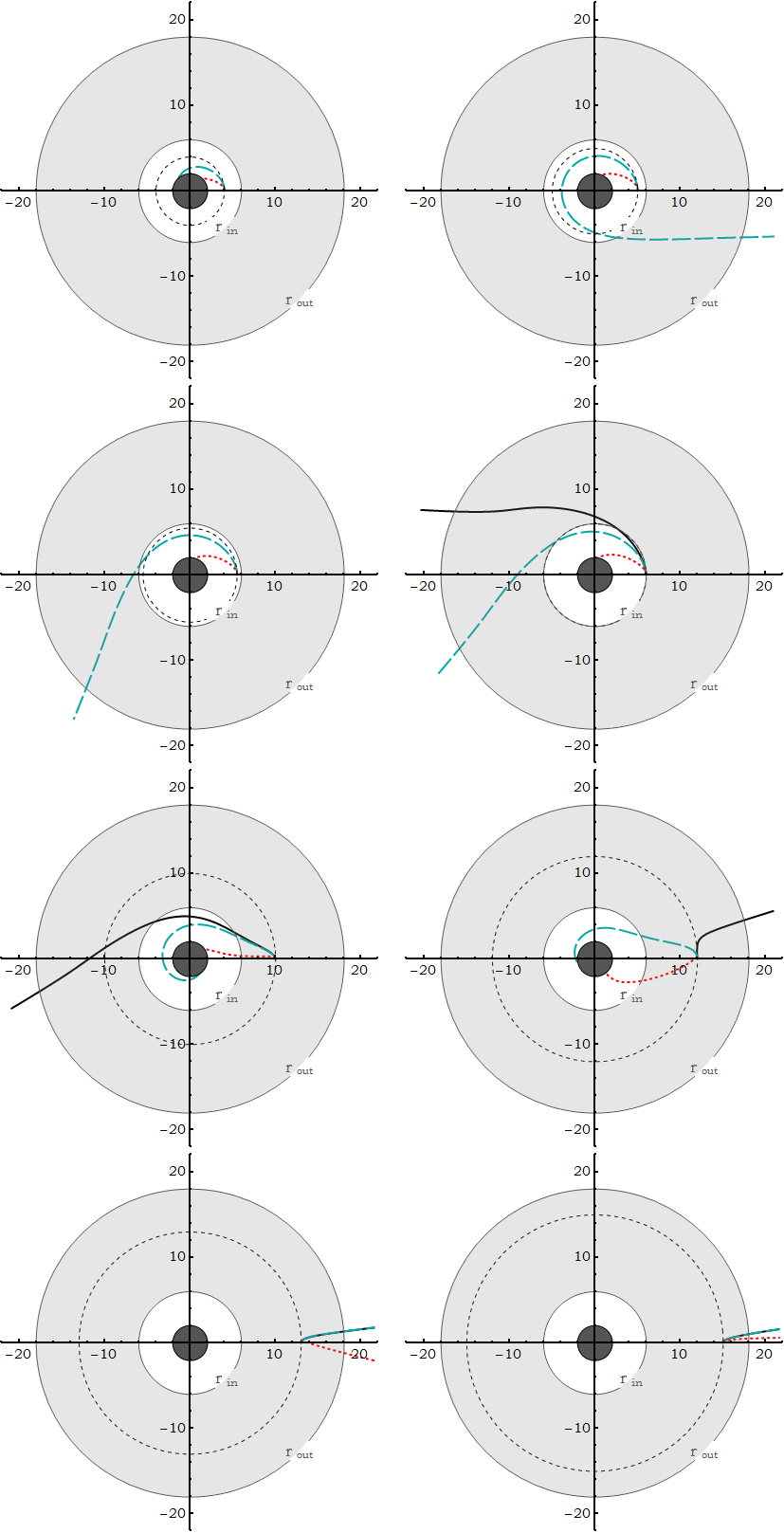}
      \caption{\texttt{Burning Arrow} results: degradation of equatorial orbits around a nonrotating BH due to the semi-opaque (LFM, Sect. 3.2d) disk's thermal radiation. With solid black is the \cite{B72} velocity profile, with dashed cyan and with dotted red an approximation of a SANE and a MAD model velocity profile respectively \citep{NSPK,PSKN,NIA}.
      }
         \label{Burning Arrow short}
   \end{figure}

The above equation gives later on, rise to the equivalent of Newton's second law of motion. It also gives us a way to calculate the accelerations present in a problem, such as those generated by the radiation forces in our case. As in the \verb+Infinity+ code, we consider different cases of velocity behaviors, such as the SANE and MAD models as mentioned before, in addition to the typical \citealt{B72} angular velocity profile.

In the beginning of the execution and depending on the study requested, the code looks up into the \verb+Infinity+ code result files and reads the appropriate ones, relevant to each case. It then uses these results to generate proper functions that give the radiation stress – energy tensor at every point of spacetime used in the problem at hand. This gives the four-acceleration due to radiation $a_{\rm rad}^{\mu }$ at each point from the extension of the aforementioned equation as:
\begin{equation}
    \frac{{{{\rm d}}^{2}}{{x}^{\mu }}}{{\rm d}{{\tau }^{2}}}+\Gamma _{\alpha \beta }^{\mu } {{u}^{\alpha }}{{u}^{\beta }}=a_{\rm rad}^{\mu } .
\end{equation}
Since the radiation acceleration is known in spacetime from the previous results, we can then solve these eight differential equations and get the four-velocity and four-positions describing the sought particle trajectory. 

The \verb+Burning Arrow+ output for degenerating orbits can be seen in Fig. \ref{Burning Arrow short}. Without the presence of radiation, the material electrons would continue moving in circular orbits, shown in the figure with the dashed circles. The presence of radiation, nevertheless, causes the electron orbits to destabilize. The electrons then, either fall into the central object or leave the AD system (gray annulus) and escape to infinity. We show here some of the results of the radiation effects on three different velocity profiles. In the following results part of this work, we examine and discuss thoroughly the noteworthy and interesting repercussions of these interactions.

\section{Discussion}
\label{S6}

In this work we presented the physics and the setup required in order to examine the radiation field produced by the hot accretion disk orbiting an astrophysical black hole. We subsequently described the five families of codes written to study these objects and investigate some of their facets. These codes study photon as well as massive particle trajectories and examine the dynamics 
of the black hole and accretion disk systems from close up and from infinity. Additionally, the codes create inclination - spin templates for assorted disk models in an attempt to assess the black hole's spin from observational data. In more, the codes study particle trajectories under the influence of relativistic effects and the presence of the disk's thermal radiation.

We looked into optically thin and thick, and geometrically thin and thick accretion disks. Also, we chose to apply certain radial and azimuthal velocity and temperature profiles that follow and agree with the majority of studies in environments such as the ones we study here. Various alterations and reassessments can be made nonetheless, in order to better suit other situations, other disk models, prograde or retrograde material motion, spacetime environments etc.

In total, for this task we ran more than 8 thousand simulations and we studied more than 4 billion photon trajectories. In order to examine and confirm the validity of our results, we corroborated that our codes successfully calculate analytically described or assorted expected trajectories for massless and massive particle, such as those explained in \citet{B72} and \citet{LPG}. Additionally, we showed that we correctly obtain trajectories of free photons moving in strong gravity environments, such as those appearing in \citet{Bini15}. In more, we adapted our codes to look toward central sources of photons, namely stars, instead of accretion tori in order to compare and cross-check our results with other, relevant studies including \citet{AEL90,ML93,ML96,LM95}, with which we found agreement for observational and physical quantities. Finally, we saw in brief here and will see in more detail further on, that our codes create observation-like images of various accretion disks at infinity which are in agreement with other similar works, such as \citet{FW04,FW07,YWF}.

The radiation field in such environments proves to be very complex for many reasons, some of which are the general relativistic effects present, and also the fact that the radiation source is no longer central, but instead extended in a large azimuthal and poloidal volume of space. In the second part of his work \citep{KpartII}, we look into the results of our codes and discuss photon trajectories and bundles, disk images and induced radiation forces, photographs of systems from infinity, their usage to estimate a black hole spin and particle orbit degeneration due to radiation. We then assess how we can use this information to draw conclusions about the physics and phenomena of these systems, their dynamical condition, their equilibrium and perhaps their evolution.

\bigskip

\begin{acknowledgements}
This work was in part supported by the General Secretariat for Research and Technology of Greece and the European Social Fund in the framework of Action “Excellence". Part of this work was performed at the Research Center for Astronomy and Applied Mathematics of the Academy of Athens.
\newline
\newline
The algorithms discussed in this work can be found at the following addresses:
\newline
\verb+Omega+: \url{https://gitlab.com/leelamichaels/Omega.git}
\newline
\verb+Infinity+: \url{https://gitlab.com/leelamichaels/Infinity.git}
\newline
\verb+Elysium+: \url{https://gitlab.com/leelamichaels/Elysium.git}
\newline
\verb+Tranquillity+: \url{https://gitlab.com/leelamichaels/Tranquillity.git}
\newline
\verb+Burning Arrow+: \url{https://gitlab.com/leelamichaels/Burning\_Arrow.git}
\end{acknowledgements}

\bibliographystyle{aa}
\typeout{}
\bibliography{AandRT.bib}

\begin{thebibliography}{124}
\expandafter\ifx\csname natexlab\endcsname\relax\def\natexlab#1{#1}\fi

\bibitem[{{Abramowicz} {et~al.}(1978){Abramowicz}, {Jaroszynski}, \&
  {Sikora}}]{A78}
{Abramowicz}, M., {Jaroszynski}, M., \& {Sikora}, M. 1978, \aap, 63, 221

\bibitem[{{Abramowicz} {et~al.}(1996){Abramowicz}, {Chen}, {Granath}, \&
  {Lasota}}]{A96}
{Abramowicz}, M.~A., {Chen}, X.~M., {Granath}, M., \& {Lasota}, J.~P. 1996,
  \apj, 471, 762

\bibitem[{{Abramowicz} {et~al.}(1988){Abramowicz}, {Czerny}, {Lasota}, \&
  {Szuszkiewicz}}]{A88}
{Abramowicz}, M.~A., {Czerny}, B., {Lasota}, J.~P., \& {Szuszkiewicz}, E. 1988,
  \apj, 332, 646

\bibitem[{{Abramowicz} {et~al.}(1990){Abramowicz}, {Ellis}, \& {Lanza}}]{AEL90}
{Abramowicz}, M.~A., {Ellis}, G. F.~R., \& {Lanza}, A. 1990, \apj, 361, 470

\bibitem[{{Abramowicz} \& {Fragile}(2013)}]{AF13}
{Abramowicz}, M.~A. \& {Fragile}, P.~C. 2013, Living Reviews in Relativity, 16,
  1

\bibitem[{{Abramowicz} {et~al.}(2010){Abramowicz}, {Jaroszy{\'n}ski}, {Kato},
  {Lasota}, {R{\'o}{\.z}a{\'n}ska}, \& {S{\k{a}}dowski}}]{A10}
{Abramowicz}, M.~A., {Jaroszy{\'n}ski}, M., {Kato}, S., {et~al.} 2010, \aap,
  521, A15

\bibitem[{{Artemova} {et~al.}(1996){Artemova}, {Bisnovatyi-Kogan},
  {Bjoernsson}, \& {Novikov}}]{Artemova}
{Artemova}, I.~V., {Bisnovatyi-Kogan}, G.~S., {Bjoernsson}, G., \& {Novikov},
  I.~D. 1996, \apj, 456, 119

\bibitem[{{Asada} {et~al.}(2016){Asada}, {Nakamura}, \& {Pu}}]{Asada16}
{Asada}, K., {Nakamura}, M., \& {Pu}, H.-Y. 2016, \apj, 833, 56

\bibitem[{{Balbus} \& {Hawley}(1991)}]{BHa91}
{Balbus}, S.~A. \& {Hawley}, J.~F. 1991, \apj, 376, 214

\bibitem[{{Balbus} \& {Hawley}(1992)}]{BHd92}
{Balbus}, S.~A. \& {Hawley}, J.~F. 1992, \apj, 400, 610

\bibitem[{{Balbus} \& {Hawley}(1998)}]{BH98}
{Balbus}, S.~A. \& {Hawley}, J.~F. 1998, Reviews of Modern Physics, 70, 1

\bibitem[{{Bardeen}(1970)}]{B70}
{Bardeen}, J.~M. 1970, \apj, 162, 71

\bibitem[{{Bardeen} {et~al.}(1972){Bardeen}, {Press}, \& {Teukolsky}}]{B72}
{Bardeen}, J.~M., {Press}, W.~H., \& {Teukolsky}, S.~A. 1972, \apj, 178, 347

\bibitem[{{Beloborodov}(1998)}]{Beloborodov98}
{Beloborodov}, A.~M. 1998, \mnras, 297, 739

\bibitem[{{Beloborodov}(1999)}]{Beloborodov99}
{Beloborodov}, A.~M. 1999, in Astronomical Society of the Pacific Conference
  Series, Vol. 161, High Energy Processes in Accreting Black Holes, ed.
  J.~{Poutanen} \& R.~{Svensson}, 295

\bibitem[{{Beloborodov}(2001)}]{Beloborodov01}
{Beloborodov}, A.~M. 2001, Advances in Space Research, 28, 411

\bibitem[{{Bildsten} \& {Rutledge}(2001)}]{Bildsten01}
{Bildsten}, L. \& {Rutledge}, R.~E. 2001, in The Neutron Star - Black Hole
  Connection, ed. C.~{Kouveliotou}, J.~{Ventura}, \& E.~{van den Heuvel}, Vol.
  567, 245

\bibitem[{{Bini} {et~al.}(2015){Bini}, {Geralico}, {Jantzen}, \&
  {Semer{\'a}k}}]{Bini15}
{Bini}, D., {Geralico}, A., {Jantzen}, R.~T., \& {Semer{\'a}k}, O. 2015,
  \mnras, 446, 2317

\bibitem[{{Bini} {et~al.}(2011){Bini}, {Geralico}, {Jantzen}, {Semer{\'a}k}, \&
  {Stella}}]{Bini11}
{Bini}, D., {Geralico}, A., {Jantzen}, R.~T., {Semer{\'a}k}, O., \& {Stella},
  L. 2011, Classical and Quantum Gravity, 28, 035008

\bibitem[{{Bini} {et~al.}(2009){Bini}, {Jantzen}, \& {Stella}}]{Bini09}
{Bini}, D., {Jantzen}, R.~T., \& {Stella}, L. 2009, Classical and Quantum
  Gravity, 26, 055009

\bibitem[{{Bisnovatyi-Kogan}(2001)}]{DynamicProcesses}
{Bisnovatyi-Kogan}, G.~S. 2001, Discrete Dynamics in Nature and Society, 6

\bibitem[{{Bisnovatyi-Kogan} {et~al.}(2002){Bisnovatyi-Kogan}, {Lovelace}, \&
  {Belinski}}]{BKLB}
{Bisnovatyi-Kogan}, G.~S., {Lovelace}, R.~V.~E., \& {Belinski}, V.~A. 2002,
  \apj, 580, 380

\bibitem[{{Blaes}(2004)}]{Blaes04}
{Blaes}, O.~M. 2004, in Accretion Discs, Jets and High Energy Phenomena in
  Astrophysics, ed. V.~{Beskin}, G.~{Henri}, F.~{Menard}, \& {et al.}, Vol.~78,
  137--185

\bibitem[{{Blandford} {et~al.}(2002){Blandford}, {Agol}, {Broderick}, {Heyl},
  {Koopmans}, \& {Lee}}]{BABHKL}
{Blandford}, R., {Agol}, E., {Broderick}, A., {et~al.} 2002, in Astrophysical
  Spectropolarimetry, ed. J.~{Trujillo-Bueno}, F.~{Moreno-Insertis}, \&
  F.~{S{\'a}nchez}, 177--223

\bibitem[{{Blandford} \& {Znajek}(1977)}]{BZ}
{Blandford}, R.~D. \& {Znajek}, R.~L. 1977, \mnras, 179, 433

\bibitem[{{Bondi}(1952)}]{Bondi1952}
{Bondi}, H. 1952, \mnras, 112, 195

\bibitem[{{Broderick} \& {Loeb}(2005)}]{BLJun05}
{Broderick}, A.~E. \& {Loeb}, A. 2005, \mnras, 363, 353

\bibitem[{{Broderick} \& {Loeb}(2006{\natexlab{a}})}]{BLSep05}
{Broderick}, A.~E. \& {Loeb}, A. 2006{\natexlab{a}}, \mnras, 367, 905

\bibitem[{{Broderick} \& {Loeb}(2006{\natexlab{b}})}]{BLJul06}
{Broderick}, A.~E. \& {Loeb}, A. 2006{\natexlab{b}}, in Journal of Physics
  Conference Series, Vol.~54, Journal of Physics Conference Series, 448--455

\bibitem[{{Carter}(1968)}]{Carter68}
{Carter}, B. 1968, Physical Review, 174, 1559

\bibitem[{{Chandrasekhar}(1983)}]{Chandra}
{Chandrasekhar}, S. 1983, {The mathematical theory of black holes}

\bibitem[{{Chatterjee} {et~al.}(2019){Chatterjee}, {Liska}, {Tchekhovskoy}, \&
  {Markoff}}]{Chatterjee19}
{Chatterjee}, K., {Liska}, M., {Tchekhovskoy}, A., \& {Markoff}, S.~B. 2019,
  \mnras, 490, 2200

\bibitem[{{Chen} \& {Podsiadlowski}(2016)}]{Chen16}
{Chen}, W.-C. \& {Podsiadlowski}, P. 2016, \apj, 830, 131

\bibitem[{{Choquet-Bruhat} {et~al.}(1977){Choquet-Bruhat}, {DeWitt-Morette}, \&
  {Dillard-Bleik}}]{AMP}
{Choquet-Bruhat}, Y., {DeWitt-Morette}, C., \& {Dillard-Bleik}, M. 1977,
  {Analysis, manifolds and physics}

\bibitem[{{Contopoulos} \& {Kazanas}(1998)}]{CB98}
{Contopoulos}, I. \& {Kazanas}, D. 1998, \apj, 508, 859

\bibitem[{{Contopoulos} {et~al.}(2006){Contopoulos}, {Kazanas}, \&
  {Christodoulou}}]{CKC06}
{Contopoulos}, I., {Kazanas}, D., \& {Christodoulou}, D.~M. 2006, \apj, 652,
  1451

\bibitem[{{Contopoulos} \& {Papadopoulos}(2012)}]{CP12}
{Contopoulos}, I. \& {Papadopoulos}, D.~B. 2012, \mnras, 425, 147

\bibitem[{{Cunningham}(1976)}]{Cunningham76}
{Cunningham}, C. 1976, \apj, 208, 534

\bibitem[{{Cunningham}(1975)}]{Cunningham75}
{Cunningham}, C.~T. 1975, \apj, 202, 788

\bibitem[{{Davelaar} {et~al.}(2018){Davelaar}, {Bronzwaer}, {Kok}, {Younsi},
  {Mo{\'s}cibrodzka}, \& {Falcke}}]{Davelaar18}
{Davelaar}, J., {Bronzwaer}, T., {Kok}, D., {et~al.} 2018, Computational
  Astrophysics and Cosmology, 5, 1

\bibitem[{{Done} {et~al.}(2007){Done}, {Gierli{\'n}ski}, \& {Kubota}}]{DGK}
{Done}, C., {Gierli{\'n}ski}, M., \& {Kubota}, A. 2007, \aapr, 15, 1

\bibitem[{{Dubus}(2003)}]{Dubus}
{Dubus}, G. 2003, in EAS Publications Series, Vol.~7, EAS Publications Series,
  ed. C.~{Motch} \& J.-M. {Hameury}, 283

\bibitem[{{Esin} {et~al.}(1997){Esin}, {McClintock}, \& {Narayan}}]{Esin97}
{Esin}, A.~A., {McClintock}, J.~E., \& {Narayan}, R. 1997, \apj, 489, 865

\bibitem[{{Event Horizon Telescope Collaboration IV}(2019)}]{EHT}
{Event Horizon Telescope Collaboration IV}. 2019, \apjl, 875, L4

\bibitem[{{Fender}(2002)}]{Fender02}
{Fender}, R. 2002, {Relativistic Outflows from X-ray Binaries
  ('Microquasars')}, ed. A.~W. {Guthmann}, M.~{Georganopoulos}, A.~{Marcowith},
  \& K.~{Manolakou}, Vol. 589, 101

\bibitem[{{Fuerst}(2006)}]{FuerstPhD}
{Fuerst}, S.~V. 2006, PhD thesis, Mullard Space Science Laboratory, University
  College London, Holmbury St. Mary, Dorking, Surrey RH5 6NT, UK

\bibitem[{{Fuerst} \& {Wu}(2004)}]{FW04}
{Fuerst}, S.~V. \& {Wu}, K. 2004, \aap, 424, 733

\bibitem[{{Fuerst} \& {Wu}(2007)}]{FW07}
{Fuerst}, S.~V. \& {Wu}, K. 2007, \aap, 474, 55

\bibitem[{{Haggard} {et~al.}(2004){Haggard}, {Cool}, {Anderson}, {Edmonds},
  {Callanan}, {Heinke}, {Grindlay}, \& {Bailyn}}]{Haggard04}
{Haggard}, D., {Cool}, A.~M., {Anderson}, J., {et~al.} 2004, \apj, 613, 512

\bibitem[{{Hawley} \& {Balbus}(1991)}]{BHb91}
{Hawley}, J.~F. \& {Balbus}, S.~A. 1991, \apj, 376, 223

\bibitem[{{Hawley} \& {Balbus}(1992)}]{BHc92}
{Hawley}, J.~F. \& {Balbus}, S.~A. 1992, \apj, 400, 595

\bibitem[{{Hawley} {et~al.}(1995){Hawley}, {Gammie}, \& {Balbus}}]{HGB}
{Hawley}, J.~F., {Gammie}, C.~F., \& {Balbus}, S.~A. 1995, \apj, 440, 742

\bibitem[{{Heinke} {et~al.}(2013){Heinke}, {Ivanova}, {Engel}, {Pavlovskii},
  {Sivakoff}, {Cartwright}, \& {Gladstone}}]{Heinke13}
{Heinke}, C.~O., {Ivanova}, N., {Engel}, M.~C., {et~al.} 2013, \apj, 768, 184

\bibitem[{{Igumenshchev} {et~al.}(2003){Igumenshchev}, {Narayan}, \&
  {Abramowicz}}]{Igumenshchev}
{Igumenshchev}, I.~V., {Narayan}, R., \& {Abramowicz}, M.~A. 2003, \apj, 592,
  1042

\bibitem[{{Inoue} \& {Hoshi}(1987)}]{InoueHoshi}
{Inoue}, H. \& {Hoshi}, R. 1987, \apj, 322, 320

\bibitem[{{Komissarov} \& {Porth}(2021)}]{Komissarov2021}
{Komissarov}, S. \& {Porth}, O. 2021, \nar, 92, 101610

\bibitem[{{Komissarov}(1999)}]{Komissarov1999}
{Komissarov}, S.~S. 1999, \mnras, 308, 1069

\bibitem[{{Komissarov}(2001)}]{Komissarov2001}
{Komissarov}, S.~S. 2001, \mnras, 326, L41

\bibitem[{{Komissarov} {et~al.}(2007){Komissarov}, {Barkov}, {Vlahakis}, \&
  {K{\"o}nigl}}]{KBVK}
{Komissarov}, S.~S., {Barkov}, M.~V., {Vlahakis}, N., \& {K{\"o}nigl}, A. 2007,
  \mnras, 380, 51

\bibitem[{{Konoplya} {et~al.}(2021){Konoplya}, {Kunz}, \&
  {Zhidenko}}]{Konoplya2021}
{Konoplya}, R.~A., {Kunz}, J., \& {Zhidenko}, A. 2021, arXiv e-prints,
  arXiv:2102.10649

\bibitem[{Koutsantoniou(2014)}]{Leela2014}
Koutsantoniou, L.~E. 2014, Master's thesis, Department of Astrophysics,
  Astronomy and Mechanics, Faculty of Physics, University of Athens,
  Panepistimiopolis Zografos, Athens 15784, Greece

\bibitem[{{Koutsantoniou}(2021b, in preparation)}]{KpartII}
{Koutsantoniou}, L.~E. 2021b, in preparation, \aap

\bibitem[{{Koutsantoniou} \& {Contopoulos}(2014)}]{KC2014}
{Koutsantoniou}, L.~E. \& {Contopoulos}, I. 2014, \apj, 794, 27

\bibitem[{{Kozlowski} {et~al.}(1978){Kozlowski}, {Jaroszynski}, \&
  {Abramowicz}}]{K78}
{Kozlowski}, M., {Jaroszynski}, M., \& {Abramowicz}, M.~A. 1978, \aap, 63, 209

\bibitem[{{Krolik}(1999{\natexlab{a}})}]{KrolikB}
{Krolik}, J.~H. 1999{\natexlab{a}}, {Active galactic nuclei : from the central
  black hole to the galactic environment}

\bibitem[{{Krolik}(1999{\natexlab{b}})}]{KrolikP}
{Krolik}, J.~H. 1999{\natexlab{b}}, in Astronomical Society of the Pacific
  Conference Series, Vol. 161, High Energy Processes in Accreting Black Holes,
  ed. J.~{Poutanen} \& R.~{Svensson}, 315

\bibitem[{{Kylafis} {et~al.}(2012){Kylafis}, {Contopoulos}, {Kazanas}, \&
  {Christodoulou}}]{KCKC}
{Kylafis}, N.~D., {Contopoulos}, I., {Kazanas}, D., \& {Christodoulou}, D.~M.
  2012, \aap, 538, A5

\bibitem[{{Lamb} \& {Miller}(1995)}]{LM95}
{Lamb}, F.~K. \& {Miller}, M.~C. 1995, \apj, 439, 828

\bibitem[{{Lasota}(1999)}]{Lasota99}
{Lasota}, J.~P. 1999, \physrep, 311, 247

\bibitem[{{Lee} {et~al.}(2000){Lee}, {Wijers}, \& {Brown}}]{Lee2000}
{Lee}, H.~K., {Wijers}, R.~A.~M.~J., \& {Brown}, G.~E. 2000, \physrep, 325, 83

\bibitem[{{Levin} \& {Grossman}(2009)}]{LG}
{Levin}, J. \& {Grossman}, R. 2009, \prd, 79, 043016

\bibitem[{{Levin} \& {Perez-Giz}(2008)}]{LPG}
{Levin}, J. \& {Perez-Giz}, G. 2008, \prd, 77, 103005

\bibitem[{{Livio} {et~al.}(1999){Livio}, {Ogilvie}, \& {Pringle}}]{Livio99}
{Livio}, M., {Ogilvie}, G.~I., \& {Pringle}, J.~E. 1999, \apj, 512, 100

\bibitem[{{Longair}(2011)}]{Longair}
{Longair}, M.~S. 2011, {High Energy Astrophysics}

\bibitem[{{Mahlmann} {et~al.}(2020){Mahlmann}, {Levinson}, \&
  {Aloy}}]{Mahlmann20}
{Mahlmann}, J.~F., {Levinson}, A., \& {Aloy}, M.~A. 2020, \mnras, 494, 4203

\bibitem[{{McKinney}(2005)}]{McKinney2005}
{McKinney}, J.~C. 2005, \apjl, 630, L5

\bibitem[{{McKinney} {et~al.}(2012){McKinney}, {Tchekhovskoy}, \&
  {Blandford}}]{MTB}
{McKinney}, J.~C., {Tchekhovskoy}, A., \& {Blandford}, R.~D. 2012, \mnras, 423,
  3083

\bibitem[{{Meyer-Hofmeister} {et~al.}(2009){Meyer-Hofmeister}, {Liu}, \&
  {Meyer}}]{Meyer09}
{Meyer-Hofmeister}, E., {Liu}, B.~F., \& {Meyer}, F. 2009, \aap, 508, 329

\bibitem[{{Miller} \& {Lamb}(1993)}]{ML93}
{Miller}, M.~C. \& {Lamb}, F.~K. 1993, \apjl, 413, L43

\bibitem[{{Miller} \& {Lamb}(1996)}]{ML96}
{Miller}, M.~C. \& {Lamb}, F.~K. 1996, \apj, 470, 1033

\bibitem[{{Misner} {et~al.}(1973){Misner}, {Thorne}, \& {Wheeler}}]{MTW}
{Misner}, C.~W., {Thorne}, K.~S., \& {Wheeler}, J.~A. 1973, {Gravitation}

\bibitem[{{Mo{\'s}cibrodzka} {et~al.}(2018){Mo{\'s}cibrodzka}, {Falcke}, \&
  {Noble}}]{Moscibrodzka18}
{Mo{\'s}cibrodzka}, M., {Falcke}, H., \& {Noble}, S. 2018, in Fourteenth Marcel
  Grossmann Meeting - MG14, ed. M.~{Bianchi}, R.~T. {Jansen}, \& R.~{Ruffini},
  3519--3524

\bibitem[{{Mo{\'s}cibrodzka} {et~al.}(2014){Mo{\'s}cibrodzka}, {Falcke},
  {Shiokawa}, \& {Gammie}}]{Moscibrodzka14}
{Mo{\'s}cibrodzka}, M., {Falcke}, H., {Shiokawa}, H., \& {Gammie}, C.~F. 2014,
  \aap, 570, A7

\bibitem[{{Mo{\'s}cibrodzka} {et~al.}(2009){Mo{\'s}cibrodzka}, {Gammie},
  {Dolence}, {Shiokawa}, \& {Leung}}]{Moscibrodzka09}
{Mo{\'s}cibrodzka}, M., {Gammie}, C.~F., {Dolence}, J.~C., {Shiokawa}, H., \&
  {Leung}, P.~K. 2009, \apj, 706, 497

\bibitem[{{Mueller} \& {Grave}(2009)}]{CatalogueofSpacetimes}
{Mueller}, T. \& {Grave}, F. 2009, arXiv e-prints, arXiv:0904.4184

\bibitem[{Müller(2004)}]{Mueller04}
Müller, A. 2004, Ph.D. Thesis: Black hole astrophysics: Magnetohydrodynamics
  on the Kerr geometry, Landessternwarte Heidelberg, Germany

\bibitem[{{Nakamura} {et~al.}(2018){Nakamura}, {Asada}, {Hada}, {Pu}, {Noble},
  {Tseng}, {Toma}, {Kino}, {Nagai}, {Takahashi}, {Algaba}, {Orienti},
  {Akiyama}, {Doi}, {Giovannini}, {Giroletti}, {Honma}, {Koyama}, {Lico},
  {Niinuma}, \& {Tazaki}}]{Nakamura18}
{Nakamura}, M., {Asada}, K., {Hada}, K., {et~al.} 2018, \apj, 868, 146

\bibitem[{{Narayan} {et~al.}(2003){Narayan}, {Igumenshchev}, \&
  {Abramowicz}}]{NIA}
{Narayan}, R., {Igumenshchev}, I.~V., \& {Abramowicz}, M.~A. 2003, \pasj, 55,
  L69

\bibitem[{{Narayan} \& {McClintock}(2008)}]{N08}
{Narayan}, R. \& {McClintock}, J.~E. 2008, \nar, 51, 733

\bibitem[{{Narayan} {et~al.}(2012){Narayan}, {S{\k{a}}dowski}, {Penna}, \&
  {Kulkarni}}]{NSPK}
{Narayan}, R., {S{\k{a}}dowski}, A., {Penna}, R.~F., \& {Kulkarni}, A.~K. 2012,
  \mnras, 426, 3241

\bibitem[{{Narayan} \& {Yi}(1994)}]{NY94}
{Narayan}, R. \& {Yi}, I. 1994, \apjl, 428, L13

\bibitem[{{Narayan} \& {Yi}(1995)}]{NY95}
{Narayan}, R. \& {Yi}, I. 1995, \apj, 444, 231

\bibitem[{{Noble} {et~al.}(2011){Noble}, {Krolik}, {Schnittman}, \&
  {Hawley}}]{Noble11}
{Noble}, S.~C., {Krolik}, J.~H., {Schnittman}, J.~D., \& {Hawley}, J.~F. 2011,
  \apj, 743, 115

\bibitem[{{Noble} {et~al.}(2007){Noble}, {Leung}, {Gammie}, \&
  {Book}}]{Noble07}
{Noble}, S.~C., {Leung}, P.~K., {Gammie}, C.~F., \& {Book}, L.~G. 2007,
  Classical and Quantum Gravity, 24, S259

\bibitem[{{Novikov} \& {Thorne}(1973)}]{Novikov-Thorne}
{Novikov}, I.~D. \& {Thorne}, K.~S. 1973, in Black Holes (Les Astres Occlus),
  343--450

\bibitem[{{Parfrey} {et~al.}(2019){Parfrey}, {Philippov}, \&
  {Cerutti}}]{Parfrey19}
{Parfrey}, K., {Philippov}, A., \& {Cerutti}, B. 2019, \prl, 122, 035101

\bibitem[{{Park} {et~al.}(2019){Park}, {Hada}, {Kino}, {Nakamura}, {Ro}, \&
  {Trippe}}]{Park19}
{Park}, J., {Hada}, K., {Kino}, M., {et~al.} 2019, \apj, 871, 257

\bibitem[{{Penna} {et~al.}(2013{\natexlab{a}}){Penna}, {Narayan}, \&
  {S{\k{a}}dowski}}]{PNS}
{Penna}, R.~F., {Narayan}, R., \& {S{\k{a}}dowski}, A. 2013{\natexlab{a}},
  \mnras, 436, 3741

\bibitem[{{Penna} {et~al.}(2013{\natexlab{b}}){Penna}, {S{\k{a}}dowski},
  {Kulkarni}, \& {Narayan}}]{PSKN}
{Penna}, R.~F., {S{\k{a}}dowski}, A., {Kulkarni}, A.~K., \& {Narayan}, R.
  2013{\natexlab{b}}, \mnras, 428, 2255

\bibitem[{{Penrose} \& {Floyd}(1971)}]{Penrose71}
{Penrose}, R. \& {Floyd}, R.~M. 1971, Nature Physical Science, 229, 177

\bibitem[{{Pessah} {et~al.}(2007){Pessah}, {Chan}, \& {Psaltis}}]{PCP}
{Pessah}, M.~E., {Chan}, C.-k., \& {Psaltis}, D. 2007, \apjl, 668, L51

\bibitem[{{Podsiadlowski} {et~al.}(2002){Podsiadlowski}, {Rappaport}, \&
  {Pfahl}}]{Pods02}
{Podsiadlowski}, P., {Rappaport}, S., \& {Pfahl}, E.~D. 2002, \apj, 565, 1107

\bibitem[{{Poynting}(1903)}]{Poynting}
{Poynting}, J.~H. 1903, \mnras, 64, A1

\bibitem[{{Quataert}(2001)}]{LREF}
{Quataert}, E. 2001, in Astronomical Society of the Pacific Conference Series,
  Vol. 224, Probing the Physics of Active Galactic Nuclei, ed. B.~M.
  {Peterson}, R.~W. {Pogge}, \& R.~S. {Polidan}, 71

\bibitem[{{Robertson}(1937)}]{Robertson}
{Robertson}, H.~P. 1937, \mnras, 97, 423

\bibitem[{{Rybicki} \& {Lightman}(1986)}]{RL}
{Rybicki}, G.~B. \& {Lightman}, A.~P. 1986, {Radiative Processes in
  Astrophysics} (Wiley-VCH)

\bibitem[{{Sadowski}(2011)}]{SadowskiPhD}
{Sadowski}, A. 2011, Ph.D. Thesis: Slim accretion disks around black holes,
  Nicolaus Copernicus Astronomical Center, Warsaw, Poland, arXiv:1108.0396

\bibitem[{{Shakura} \& {Sunyaev}(1973)}]{SS73}
{Shakura}, N.~I. \& {Sunyaev}, R.~A. 1973, \aap, 500, 33

\bibitem[{{S{\k{a}}dowski}(2009)}]{SadowskiApJ}
{S{\k{a}}dowski}, A. 2009, \apjs, 183, 171

\bibitem[{{S{\k{a}}dowski}(2016)}]{016b}
{S{\k{a}}dowski}, A. 2016, \mnras, 459, 4397

\bibitem[{{S{\k{a}}dowski} {et~al.}(2016){S{\k{a}}dowski}, {Lasota},
  {Abramowicz}, \& {Narayan}}]{016a}
{S{\k{a}}dowski}, A., {Lasota}, J.-P., {Abramowicz}, M.~A., \& {Narayan}, R.
  2016, \mnras, 456, 3915

\bibitem[{{Takahashi} {et~al.}(1995){Takahashi}, {Fukue}, {Sanbuichi}, \&
  {Umemura}}]{Takahashi95}
{Takahashi}, A., {Fukue}, J., {Sanbuichi}, K., \& {Umemura}, M. 1995, \pasj,
  47, 425

\bibitem[{{Tauris} \& {van den Heuvel}(2006)}]{FECS}
{Tauris}, T.~M. \& {van den Heuvel}, E.~P.~J. 2006, {Formation and evolution of
  compact stellar X-ray sources}, Vol.~39, 623--665

\bibitem[{{Tauris} {et~al.}(2000){Tauris}, {van den Heuvel}, \&
  {Savonije}}]{Tauris00}
{Tauris}, T.~M., {van den Heuvel}, E. P.~J., \& {Savonije}, G.~J. 2000, \apjl,
  530, L93

\bibitem[{{Tchekhovskoy} {et~al.}(2011){Tchekhovskoy}, {Narayan}, \&
  {McKinney}}]{TNM}
{Tchekhovskoy}, A., {Narayan}, R., \& {McKinney}, J.~C. 2011, \mnras, 418, L79

\bibitem[{{Thorne} \& {Price}(1975)}]{TP75}
{Thorne}, K.~S. \& {Price}, R.~H. 1975, \apjl, 195, L101

\bibitem[{{Turner} {et~al.}(2003){Turner}, {Stone}, {Krolik}, \& {Sano}}]{TSKS}
{Turner}, N.~J., {Stone}, J.~M., {Krolik}, J.~H., \& {Sano}, T. 2003, \apj,
  593, 992

\bibitem[{{van Haaften} {et~al.}(2012){van Haaften}, {Nelemans}, {Voss},
  {Wood}, \& {Kuijpers}}]{vanHaaften12}
{van Haaften}, L.~M., {Nelemans}, G., {Voss}, R., {Wood}, M.~A., \& {Kuijpers},
  J. 2012, \aap, 537, A104

\bibitem[{{Verbunt}(1999)}]{Verbunt99}
{Verbunt}, F. 1999, in Astronomical Society of the Pacific Conference Series,
  Vol. 160, Astrophysical Discs - an EC Summer School, ed. J.~A. {Sellwood} \&
  J.~{Goodman}, 21

\bibitem[{{Vlahakis} \& {K{\"o}nigl}(2004)}]{NB04}
{Vlahakis}, N. \& {K{\"o}nigl}, A. 2004, \apj, 605, 656

\bibitem[{{Wheeler}(2004)}]{Wheeler04}
{Wheeler}, C.~J. 2004, Advances in Space Research, 34, 2744, new X-Ray Results,
  the Next Generation of X-Ray Observatories and Gamma Ray Burst Afterglow
  Physics

\bibitem[{{Wilkins}(1972)}]{Wilkins72}
{Wilkins}, D.~C. 1972, \prd, 5, 814

\bibitem[{{Younsi} {et~al.}(2012){Younsi}, {Wu}, \& {Fuerst}}]{YWF}
{Younsi}, Z., {Wu}, K., \& {Fuerst}, S.~V. 2012, \aap, 545, A13

\bibitem[{{Yuan} {et~al.}(2019){Yuan}, {Blandford}, \& {Wilkins}}]{Yuan19}
{Yuan}, Y., {Blandford}, R.~D., \& {Wilkins}, D.~R. 2019, \mnras, 484, 4920

\end{thebibliography}

\end{document}